\documentclass[superscriptaddress,rmp,12pt,aps]{revtex4}

\usepackage{amsmath}
\usepackage{amssymb}
\usepackage{amsfonts}
\usepackage{mathrsfs}
\usepackage{graphicx}
\usepackage[usenames]{color}
\definecolor{mygrey}{gray}{0.75}

\begin{document}

\title{L{\'e}vy Statistics and Anomalous Transport:
L{\'e}vy flights and Subdiffusion}

\author{\textsc{Ralf Metzler}}
\affiliation{Physics Department, Technical University of Munich,
James Franck Stra{\ss}e, 85747 Garching, Germany}
\email{metz@ph.tum.de}
\author{\textsc{Aleksei V. Chechkin}}
\affiliation{Institute for Theoretical Physics NSC KIPT, Akademicheskaya
st.1, 61108 Kharkov, Ukraine}
\author{\textsc{Joseph Klafter}}
\affiliation{School of Chemistry, Tel Aviv University, 69978 Tel Aviv Israel}

\maketitle

\tableofcontents

\section*{Glossary}

\paragraph*{Ageing.} A dynamical process involving a long-tailed waiting time
distribution $\psi(t)\sim\tau^{\alpha}t^{-1-\alpha}$ ($0<\alpha<1$) does not
possess a characteristic time scale $\int_0^{\infty}t\psi(t)dt$ separating
microscopic and macroscopic times. Instead, such a process exhibits distinct
memory, such that the now-state of the system is strongly influenced by its
state in the past. When the system is prepared at some time $t_0<0$ and the
measurement starts at $t=0$, this means that the result of the measurement
depends on the time $t_0$ even until rather long times.\\

\paragraph*{Anomalous diffusion.} Under anomalous diffusion we understand
here deviations of the linear time dependence $\langle x^2(t)\rangle=2Kt$,
of the mean squared displacement in absence of an external bias, in the
form of a power-law: $\langle x^2(t)\rangle=2K_{\zeta}t^{\zeta}$. Here,
$K_{\zeta}$ is the anomalous diffusion constant of dimension $\mathrm{cm}^2/
\mathrm{sec}^{\zeta}$. In the range $0<\zeta<1$, we deal with
subdiffusion,
whereas $\zeta>1$ describes superdiffusion. L{\'e}vy flights have a 
diverging mean squared displacement.\\

\paragraph*{Continuous time random walk.} Continuous time random walk theory
describes a random motion by assigning each jump a jump length $x$ and a
waiting time $t$ elapsing in between two successive jumps, drawn from the
two probability densities $\lambda(x)$
and $\psi(t)$, respectively. The two densities $\lambda$ and $\psi$ fully
specify the probability density function $P(x,t)$ describing the random
process. In Fourier-Laplace space, the propagator follows as $P(k,u)=
u^{-1}\psi(u)/[1-\psi(u)\lambda(k)]$.\\

\paragraph*{Fourier and Laplace transforms.} Linear partial differential
equations are often conveniently solved using integral transforms of the
Laplace and Fourier type. Moreover, the definitions of the fractional
operators used in the text correspond to Laplace and Fourier convolutions,
such that these transformations also become useful there. The Laplace
transform of a function $f(t)$ is defined as
\[
f(u)\equiv\mathscr{L}\left\{f(t);u\right\}=\int_0^{\infty}f(t)e^{-ut}dt,
\]
while the Fourier transform of $g(x)$ reads
\[
g(k)\equiv\mathscr{F}\left\{g(x);k\right\}=\int_{-\infty}^{\infty}g(x)
e^{ikx}dx.
\]
Note that we denote the transform of a function by explicit dependence on
the respective variable. For $u=0$ and $k=0$ the Laplace and Fourier
transform is but the average of the function $f$ or $g$, respectively.
Tauberian theorems ascertain relations between the original function and
its transform. For instance, the small $u$ behaviour $f(u)\sim 1-(u\tau)^
{\alpha}$ implies the long time scaling $f(t)\sim\tau^{\alpha}/t^{1+\alpha}$.
See Refs.~\cite{hughes,feller} for details.\\

\paragraph*{Fractional differintegration.} The multiple derivative of an
integer power is $d^mx^n/dx^m=n!/(n-m)!x^{n-m}$ for $m\ge n$. The result is
zero if $m>n$. Replacing the factorials by the $\Gamma$-function, one
can generalise this relation to $_0D_x^qx^p=\Gamma(1+p)t^{p-q}/\Gamma(1+p-q)$.
In particular, this includes the fractional differentiation of a constant,
$_0D_x^q1=x^{-q}/\Gamma(1-q)$, that does no longer vanish. Fractional
differentiation was first mentioned by Leibniz in a letter to de l'Hospital
in 1695. The Riemann-Liouville fractional operator used in the following
is a similarly straightforward generalisation of the Cauchy multiple
integral, followed by regular differentiation.\\

\paragraph*{L{\'e}vy flight.} A L{\'e}vy flight is a special type of
continuous time random walk. Its waiting time distribution is narrow
for instance, Poissonian with $\psi(t)=\tau^{-1}\exp(-t/\tau)$, and
the resulting dynamics therefore Markovian. The jump length distribution
of a L{\'e}vy flight is long-tailed: $\lambda(x)\sim \sigma^{\mu}|x|^{-1-
\mu}$, with $0<\mu<2$, such that no second moment exists. The resulting
PDF is a L{\'e}vy stable law with Fourier transform $P(k,t)=\exp\left(-
K^{(\mu)}t|k|^{\mu}\right)$.\\

\paragraph*{L{\'e}vy stable laws.} The generalised central limit theorem
states that the properly normalised sum of independent, identically
distributed random variables with finite variance converges to a Gaussian
limit distribution. A generalisation of this theory exists for the case
with infinite variance, namely, the generalised central limit theorem. The
related distributions are the L{\'e}vy stable laws, whose density in the
simplest case have the characteristic function $p(k)=\int_{-\infty}^{\infty}
p(x)\exp(ikx)dx$ (Fourier transform) of the form $p(k)=\exp\left(-c|k|^{\mu}
\right)$, with $0<\mu<2$. In direct space, this corresponds to a power-law
asymptotic $p(x)\simeq c|x|^{-1-\mu}$. In the limit $\mu=2$, the universal
Gaussian distribution is recovered.\\

\paragraph*{L{\'e}vy walks.} In contrast to L{\'e}vy flights, L{\'e}vy walks
possess a finite mean squared displacement, albeit having a broad jump length
distribution. This is possible by the introduction of a time penalty for long
jumps through a coupling $\lambda(x)p(t|x)$ between waiting times and jump
lengths, such that long jumps involve a longer waiting time. For instance,
a $\delta$-coupling of the form $\frac{1}{2}\lambda(x)\delta(|x|-vt)$ is
often chosen, such that $v$ plays the role of a velocity.\\

\paragraph*{Strange kinetics.} Often, deviations from exponential relaxation
patterns $\exp(-t/\tau)$ and regular Brownian diffusion are observed. Instead,
non-exponential relaxation, for instance, of the stretched exponential form
$\exp\left(-[t/\tau]^{\alpha}\right)$ ($0<\alpha<1$) or of the inverse
power-law form $t^{-\zeta}$, are observed, or anomalous diffusion behaviour
is found.\\

\paragraph*{Weak ergodicity breaking.} In a system with a broadly distributed
waiting time with diverging characteristic time scale, a particle can get
stuck at a certain position for a long time. For instance, for a waiting time
distribution of the form $\psi(t)\sim\tau^{\alpha}t^{-1-\alpha}$, the
probability of not moving until time $t$ scales as $\tau^{\alpha}t^{-\alpha}$,
i.e., decays very slowly. The probability for not moving is therefore
appreciable even for long times. A particle governed by such a $\psi(t)$
even at stationarity does not equally explore a different domains of phase
space.

\section{Definition and importance of anomalous diffusion}

\begin{figure}
\unitlength=1cm
\begin{picture}(8,8.8)
\put(-4.5,-1.4){\includegraphics{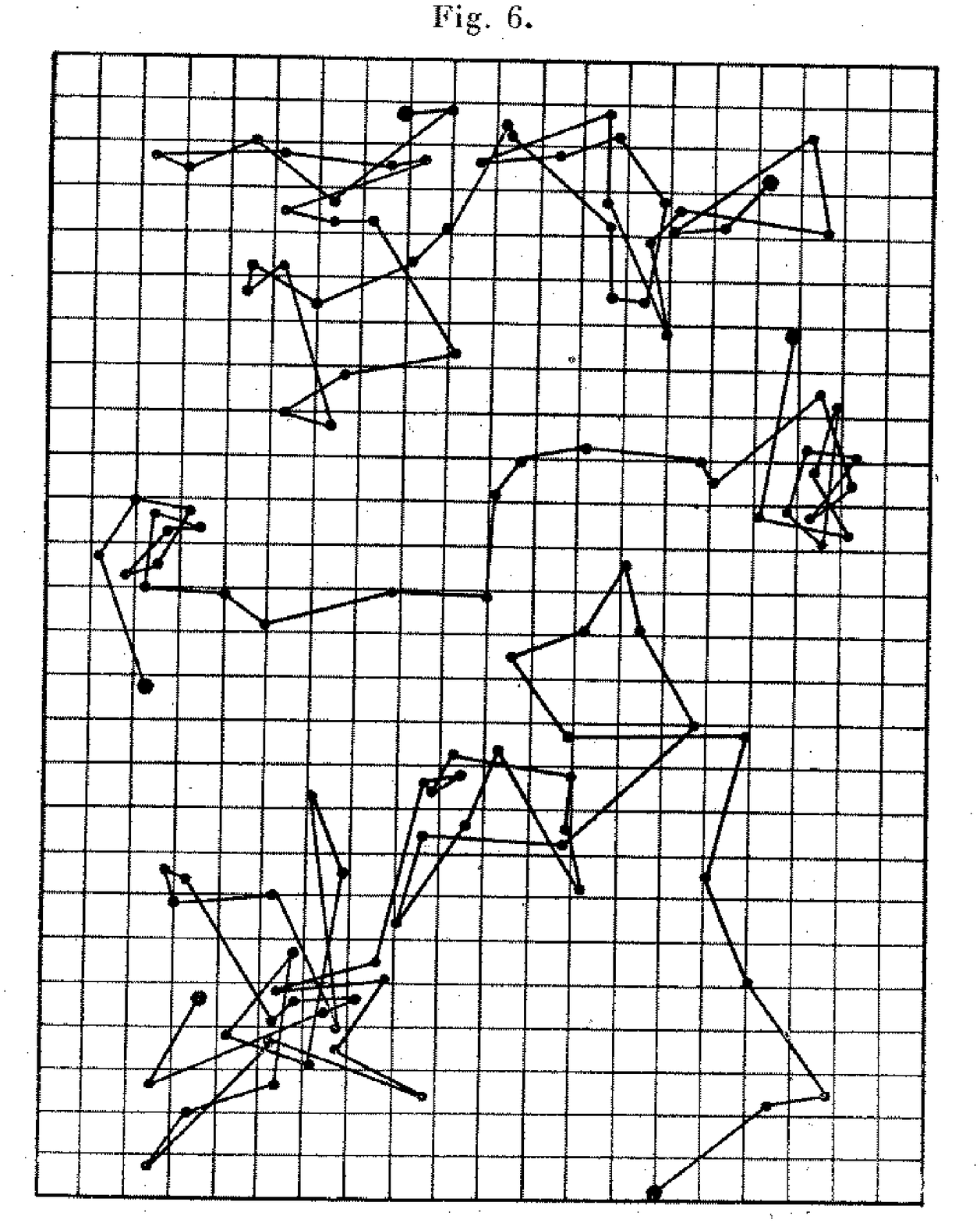}}
\put(1.8,8.0){\includegraphics{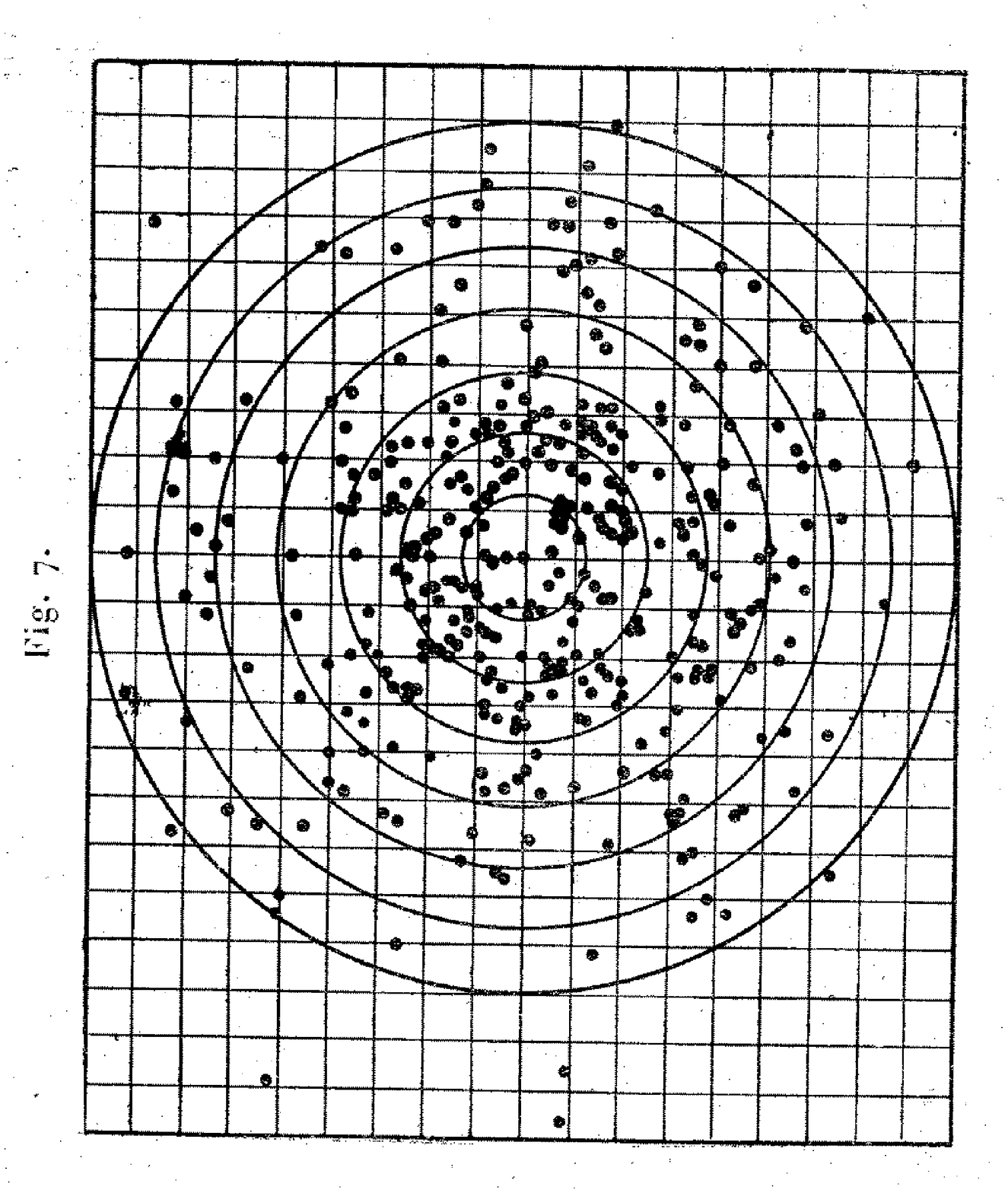}}
\end{picture}
\caption{Random walk trajectories of putty particles in water recorded by
Perrin \protect\cite{perrin1}. Left:
three designs obtained by tracing a small grain of putty at intervals of
30sec. Right: the starting point of each motion event is shifted
to the origin. The figure illustrates the continuum approach of the jump
length distribution if only a large number of jumps is considered.}
\label{perrin}
\end{figure}

\begin{figure}
\includegraphics[width=16.6cm]{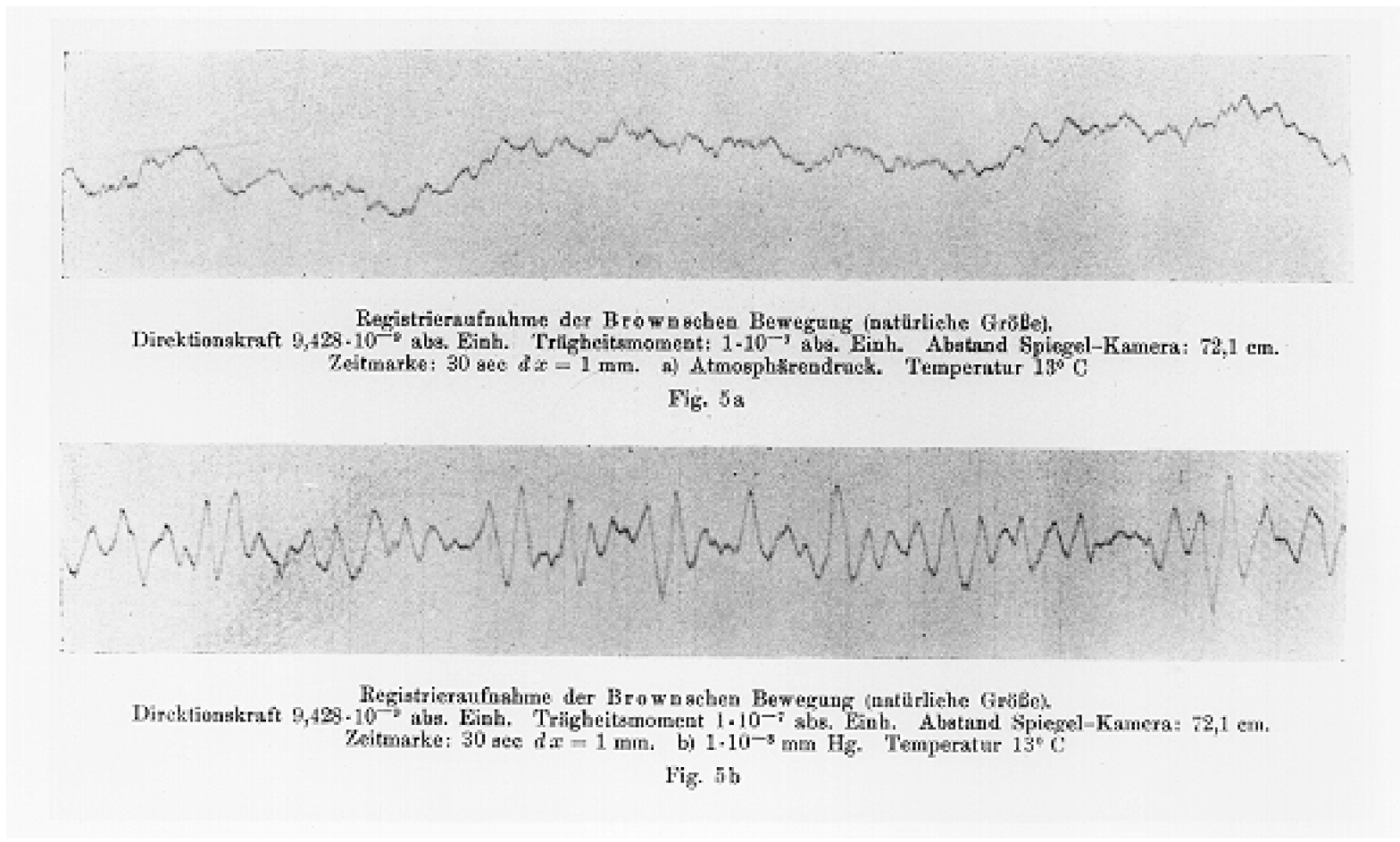}
\caption{Erroneous behaviour of Brownian motion made visible in
a high--precision measurement.
Data from an Edelmann recorder, obtained by Kappler in 1931
\protect\cite{kappler}. Kappler monitored the Brownian motion of
a small mirror (surface approx. $1{\rm mm}^2$), suspended from a fine
quartz thread (several cm long and some $\mu$m thick). The mean
squared of the torsional displacement, $\overline{\varphi^2}$,
follows the relation $D\overline{\varphi^2}=k_BT$, where $D$ is the
directional force of the suspension \protect\cite{kappler}. The
facsimiles show three different realisations. From his data,
Kappler obtained the Avogadro--Loschmidt number $N_L=60.59
\cdot 10^{22} \pm 1\%$, to a remarkable accuracy.}
\label{kappler}
\end{figure}

Classical Brownian motion characterised by a mean squared displacement
\begin{equation}
\label{msd}
\langle x^2(t)\rangle=2Kt,
\end{equation}
growing linear in time in absence of an external bias is the paradigm
for random motion. Note that we restrict our discussion to one
dimension. It quantifies the jittery motion of coal dust particles
observed by Dutchman Jan Ingenhousz in 1785 \cite{ingenhousz}, the
zigzagging of pollen grain in solution reported by Robert Brown in 1827
\cite{brown}, and possibly the dance of dust particles in the beam of
sunlight in a stairwell so beautifully embalmed in the famed poem by
Lucretius \cite{carus}. In general,
Brownian motion occurs in simple, sufficiently homogeneous systems such
as simple liquids or gasses. In the continuum limit, Brownian motion is
governed by the diffusion equation
\begin{equation}
\label{de}
\frac{\partial P(x,t)}{\partial t}=K\frac{\partial^2}{\partial x^2}P(x,t)
\end{equation}
for the probability density function (PDF) $P(x,t)$ describing the
probability $P(x,t)dx$ to find the particle at a position in the interval
$x,\ldots,x+dx$ at time $t$. For a point-like initial condition $P(x,0)=
\delta(x)$, the solution becomes the celebrated Gaussian PDF
\begin{equation}
P(x,t)=\frac{1}{\sqrt{4\pi Kt}}\exp\left(-\frac{x^2}{4Kt}\right).
\end{equation}
The diffusion constant $K$ fulfils the
Einstein-Stokes relation $K=k_BT/(m\eta)$, where $k_BT$ is the Boltzmann
energy at temperature $T$, $m$ is the mass of the test particle, and $\eta$
the friction coefficient. This relation between microscopic and macroscopic
quantities was used for the determination of the Avogadro number by Perrin
\cite{perrin}. Examples of Perrin's recorded random walks and the jump length
distribution constructed from the data for a time increment of 30 sec are
shown in Fig.~\ref{perrin}. Fig.~\ref{kappler} displays time traces of the
Brownian motion of a small mirror in air that lead to an amazingly accurate
determination of the Avogadro number by Keppler, for two different ambient
pressures \cite{kappler}.

However, in many systems deviations from the linear behaviour
(\ref{msd}) are observed \cite{bouchaudgeorges,report,report1,soklablu,%
zaslavsky,chechkinreview,physicstoday}.
These deviations can assume the power-law form
\begin{equation}
\langle x^2(t)\rangle=2K_{\zeta}t^{\zeta}.
\end{equation}
For $0<\zeta<1$, we observe subdiffusion. Prominent examples of subdiffusion
include charge carrier transport in amorphous semiconductors
\cite{pfister,schermo}, diffusion of chemicals in subsurface aquifers
\cite{grl}, the motion of beads in actin gels \cite{weitz}, motion
in chaotic maps \cite{zuklan,eli},
or the subdiffusion of biomacromolecules in cells \cite{golding};
just to name a few [compare \cite{report1} for more details]. Subdiffusion of
this type is characterised by a long-tailed waiting time PDF $\psi(t)\simeq
t^{-1-\alpha}$, corresponding to the time-fractional diffusion equation
\begin{equation}
\frac{\partial P(x,t)}{\partial t}=K_{\alpha}\,_0D_t^{1-\alpha}\frac{
\partial^2}{\partial x^2}P(x,t),
\end{equation}
with the fractional Riemann-Liouville operator \cite{samko,oldham,podlubny}
\begin{equation}
_0D_t^{1-\alpha}P(x,t)=\frac{1}{\Gamma(\alpha)}\frac{\partial}{\partial t}
\int_0^t\frac{P(x,t')}{(t-t')^{1-\alpha}}dt'
\end{equation}
From the latter definition,
it becomes apparent that subdiffusion corresponds to a slowly decaying
memory integral in the dynamical equation for $P(x,t)$.

The lack of a characteristic time scale (i.e., the divergence of
$\int_0^{\infty}t\psi(t)dt$) of the waiting time PDF $\psi(t)\simeq
t^{-1-\alpha}$
no longer permits to distinguish microscopic and macroscopic events. This
causes that the diffusing particle can get stuck at a certain position for
very long times, quantified by the sticking probability $\phi(t)=\int_t^
{\infty}\psi(t')dt'$ of not moving. The PDF $P(x,t)$ exhibits characteristic
cusps at the location where the particle was initially released and, in the
presence of an external drift, growing asymmetry. Moreover, a process whose
measurement begins at $t=0$ depends on the preparation time at some prior
time, the so-called ageing \cite{eli,bouchaudage}. Another effect in this
context is that a particle
no longer evenly distributes in a certain space, such that despite the
existence of stationary states a weak ergodicity breaking occurs
\cite{eliweb,bouchaudweb,michaelweb}.

\begin{figure}
\unitlength=1cm
\begin{picture}(12,3.6)
\put(0.2,-2.8){
\includegraphics{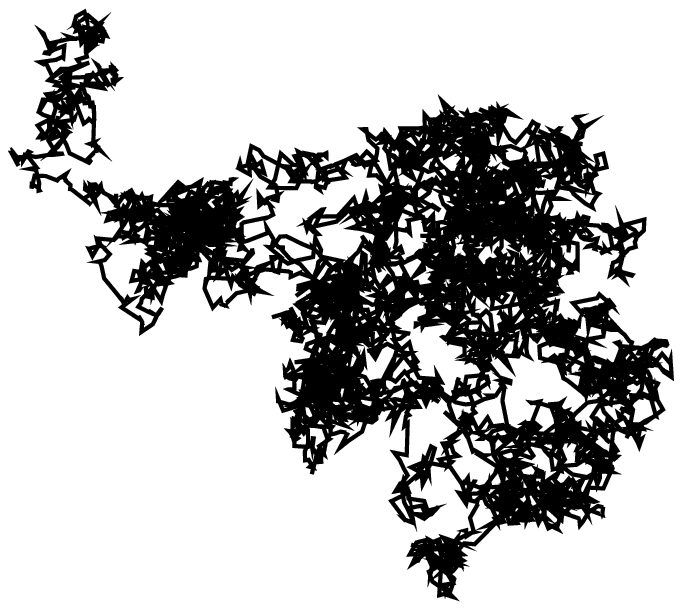}
\includegraphics{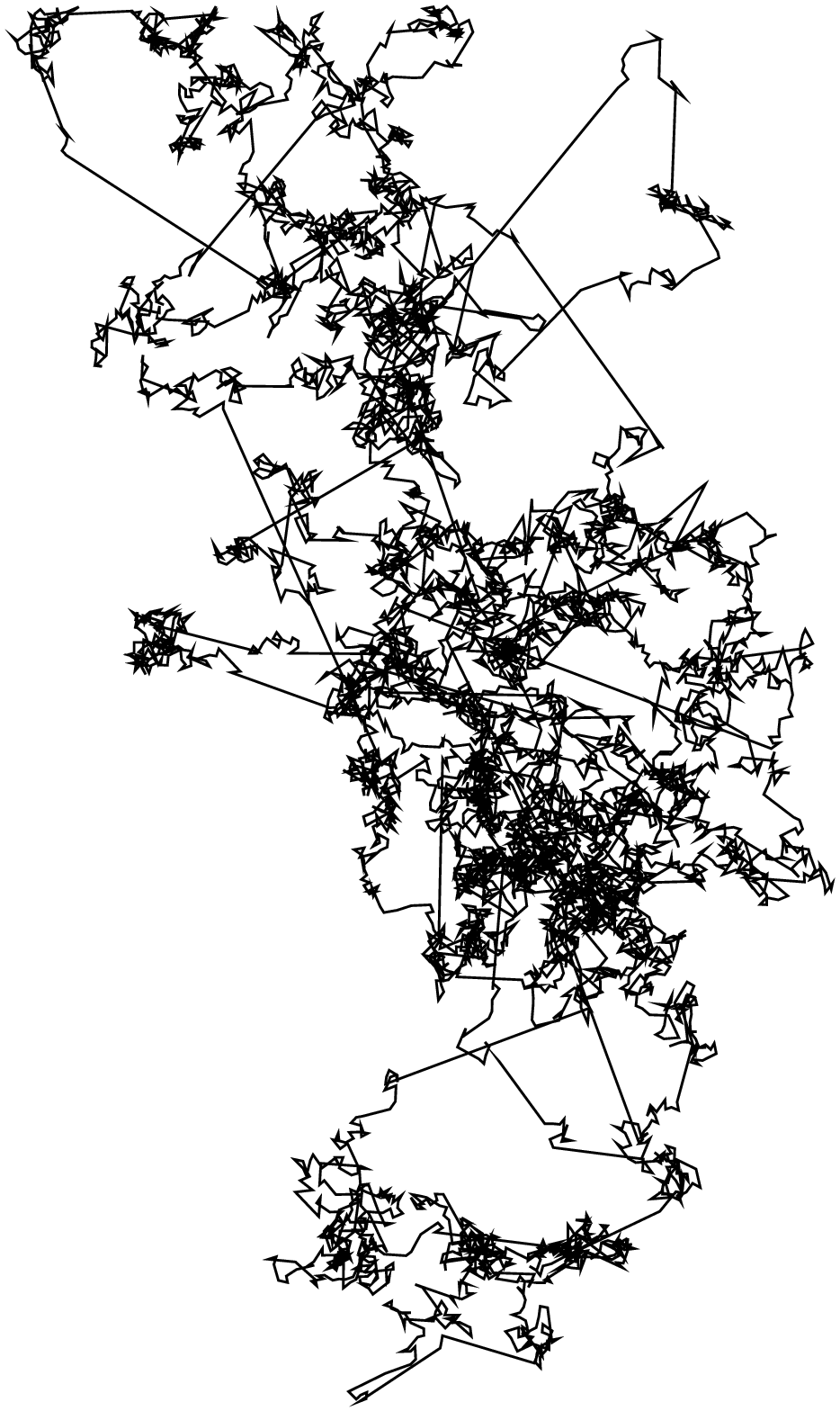}}
\end{picture}
\caption{Trajectories of Brownian motion (left) and a
L{\'e}vy flight of index $\mu=1.5$ (right), both with
the same number ($\simeq 7000$) of steps. The long sojourns and clustering
appearance of the L{\'e}vy flight are distinct.
\label{traj}}
\end{figure}

Contrasting a subdiffusing particle are L{\'e}vy flights (LFs), that are
based on a waiting time PDF with finite characteristic time but a jump
length distribution $\lambda(x)\simeq|x|^{-1-\mu}$ ($0<\mu<2$) with
diverging variance $\int_{-\infty}^{\infty}|x|^2\lambda(x)dx$. The
diffusion equation for an LF becomes generalised to the space-fractional
diffusion equation
\begin{equation}
\label{sfde}
\frac{\partial P(x,t)}{\partial t}=K^{(\mu)}\frac{\partial^{\mu}}{\partial
|x|^{\mu}}P(x,t),
\end{equation}
where the fractional Riesz-Weyl operator is defined through its Fourier
transform $\mathscr{F}\left\{\partial^{\mu}/\partial|x|^{\mu}P(x,t)\right\}
=-|k|^{\mu}P(k,t)$. In Fourier space, one therefore recovers
immediately the characteristic function $P(k,t)=\exp\left(-K^{(\mu)}|k|^{\mu}t
\right)$ of a symmetric L{\'e}vy stable law. As can be seen from
Eq.~(\ref{sfde}) LFs are Markovian processes. However, their trajectory now
has a fractal dimension $d_f=\mu$ and is characterised by local search
changing with long excursions, see Fig.~\ref{traj}. This property has been
shown to be a better
search strategy than Brownian motion, as it oversamples less
\cite{michael,stanleysearch}.
In fact, animals like albatross \cite{albatross}, spider monkeys \cite{spider},
jackals \cite{jackals}, and even plankton \cite{plankton} and bacteria
\cite{bacteria} were claimed to follow
L{\'e}vy search strategies. LFs also occur in diffusion in energy space
\cite{sms}, or in optical lattices \cite{opticallattice}. Due to
the clustering
nature of their trajectory, LFs also exhibit a form of ergodicity breaking
\cite{lutzweb}.

\section{Introduction}

For sums of independent, identically distributed random variables with
proper normalisation to the sample size, the generalised central limit
theorem guarantees the convergence of the associated probability density
to a L{\'e}vy stable density even though the variance of
these random variables diverges \cite{levy,feller,gnedenko,hughes,uchaikin}.
Well-known examples for L{\'e}vy stable densities are the one-sided
(defined for $x\ge0$) L{\'e}vy-Smirnov
distribution
\begin{equation}
\label{smirnov}
f_{1/2,-1/2}(x)=\sqrt{\frac{1}{2\pi x^3}}\exp\left(-\frac{1}{2x}\right),
\end{equation}
related to the first passage time density of a Gaussian random walk process
of passing the origin (see below), and the Cauchy (or Lorentz) distribution
\begin{equation}
\label{lorentz}
f_{1,0}(x)=\frac{1}{\pi\left(1+x^2\right)}.
\end{equation}
In general, an L{\'e}vy stable density is
defined through its characteristic function of the PDF $f(x)$
\begin{equation}
\label{character}
\varphi(z)\equiv\mathscr{F}\left\{f(x)\right\}=\int_{-\infty}^{\infty}
f_{\mu,\beta}(x)e^{ikx}dx
\end{equation}
where
\begin{equation}
\label{lsd}
\log\varphi(z)=-|z|^{\mu}\exp\left\{i\frac{\pi\beta}{2}
\mathrm{sign}(z)
\right\},
\end{equation}
for $\mu\neq 1$. Here, the skewness (or asymmetry) parameter $\beta$ is
restricted to the following region:
\begin{equation}
|\beta|\le\left\{\begin{array}{ll} \mu, & \mbox{if } 0<\mu<1\\
2-\mu, & \mbox{if } 1<\mu<2.\end{array}\right.
\end{equation}
For $\beta=0$, the corresponding L{\'e}vy stable density is symmetric around $x=0$, while
for $\beta=-\mu$ and $0<\mu<1$, it is one-sided. In general, an
L{\'e}vy stable density follows the power-law asymptotic behaviour
\begin{equation}
\label{asym}
f_{\mu,\beta}(x)\sim\frac{A_{\mu,\beta}}{|x|^{1+\mu}},\,\,\,
\mu<2,
\end{equation}
with $A_{\mu,\beta}$ being a constant, such that for all L{\'e}vy stable densities with
$\mu<2$ the variance diverges
\begin{equation}
\langle x^2\rangle=\infty.
\end{equation}
Conversely, all fractional moments $\langle|x|^{\delta}\rangle<\infty$
for all $0<\delta<\mu\le 2$. From above definitions it is obvious that
the L{\'e}vy stable density $f_{2,0}$ corresponds to the Gaussian normal distribution
\begin{equation}
f_{2,0}(x)=\sqrt{\frac{1}{4\pi}}\exp\left(-\frac{1}{4}x^2\right)
\end{equation}
possessing finite moments of any order. In this limit, the generalised
central limit theorem coincides with the more traditional, and universal,
central limit theorem.

Brownian motion has traditionally been employed as the dominant model of
choice for random noise in continuous-time systems, due to its remarkable
statistical properties and its amenability to mathematical analysis.
However, Brownian motion is just a single example of the L{\'e}vy family.
Moreover, it is a very special and somewhat misrepresenting member of this
family. Amongst the L{\'e}vy family, the Brownian member is the only motion
with continuous sample-paths. All other motions have discontinuous
trajectories, exhibiting jumps. Moreover, the L{\'e}vy family is
characterized by selfsimilar motions. Brownian motion is the only selfsimilar
L{\'e}vy motion possessing finite variance, while all other selfsimilar
L{\'e}vy motions have an infinite variance.

Random processes whose spatial coordinate $x$ or clock time $t$ are distributed
according to an L{\'e}vy stable density exhibit anomalies, that is, no longer follow the laws
of Brownian motion. Consider a continuous time random walk
process defined in terms of the jump length and waiting time distributions
$\lambda(x)$ and $\psi(t)$ \cite{montroll,schermo}. Each jump event of this
random walk, that is, is characterised by a jump length $x$ drawn from the
distribution $\lambda$, and the time $t$ between two jump events is
distributed according to $\psi$. (Note that an individual jump is supposed to
occur spontaneously.) In absence of an external bias, continuous time random
walk theory connects $\lambda(x)$ and $\psi(t)$ with the probability
distribution $P(x,t)dx$ to find the random walker at a position in the
interval $(x,x+dx)$ at time $t$. In Fourier-Laplace space, $P(k,u)\equiv
\mathscr{F}\left\{\mathscr{L}\left\{P(x,t);t\to u\right\};x\to k\right\}$, this
relation reads \cite{klablushle}
\begin{equation}
\label{pfl}
P(k,u)=\frac{1-\psi(u)}{u}\frac{1}{1-\lambda(k)\psi(u)},
\end{equation}
where $\mathscr{L}\{f(t)\}\equiv\int_0^{\infty}\exp(-ut)f(t)dt$. We here
neglect potential complications due to ageing effects. The following cases
can be distinguished:

(i) $\lambda(x)$ is Gaussian with variance $\sigma^2$ and $\psi(t)=
\delta(t-\tau)$. Then, to leading order in $k^2$ and $u$, respectively,
one obtains $\lambda(k)\simeq 1-\sigma^2k^2$ and $\psi(u)\simeq 1-u\tau$.
From relation (\ref{pfl}) one recovers the Gaussian probability density
$P(x,t)=\sqrt{1/(4\pi Kt)}\exp\{-x^2/(4Kt)\}$ with diffusion constant
$K=\sigma^2/\tau$. The corresponding mean squared displacement grows
linearly with time, see Eq.~(\ref{msd}).
This case corresponds to the continuum limit of regular Brownian motion. Note
that here and in the following, we restrict the discussion to one dimension.

(ii) Assume $\lambda(x)$ still to be Gaussian, while for the waiting time
distribution $\psi(t)$ we choose a one-sided L{\'e}vy stable density with stable index $0<
\alpha<1$. Consequently, $\psi(u)\simeq 1-(u\tau)^{\alpha}$, and the
characteristic waiting time $\int_0^{\infty}t\psi(t)dt$ diverges.
Due to this lack of a time scale separating microscopic (single jump events)
and macroscopic (on the level of $P(x,t)$) scales, $P(x,t)$ is no more
Gaussian, but given by a more complex $H$-function
\cite{schneider,report,report1}. In Fourier space, however, one finds
the quite simple analytical form \cite{report}
\begin{equation}
P(k,t)=E_{\alpha}\left(-K_{\alpha}k^2t^{\alpha}\right)=\sum_0^{\infty}
\frac{\left(K_{\alpha}k^2t^{\alpha}\right)^n}{\Gamma(1+\alpha n)}
\end{equation}
in terms of the Mittag-Leffler function. This generalised relaxation function
of the Fourier modes turns from an initial stretched exponential (KWW)
behaviour
\begin{equation}
P(k,t)\sim1-K_{\alpha}k^2t^{\alpha}/\Gamma(1+\alpha)
\sim\exp\left\{-K_{\alpha}k^2t^{\alpha}/\Gamma(1+\alpha)\right\}
\end{equation}
to a terminal power-law behaviour \cite{report}
\begin{equation}
P(k,t)\sim\Big(K_{\alpha}k^2t^{\alpha}\Gamma(1-\alpha)\Big)^{-1}.
\end{equation}
In the limit $\alpha\to 1$, it reduces to the
traditional exponential $P(k,t)=\exp(-Kk^2t)$ with finite characteristic 
waiting time. Also the mean squared
displacement changes from its linear to the power-law time dependence
\begin{equation}
\langle x^2(t)\rangle=2K_{\alpha}t^{\alpha},
\end{equation}
with $K_{\alpha}=\sigma^2/\tau_0^{\alpha}$. This is the case of
\emph{subdiffusion}. We note that in $x,t$ space the dynamical equation is
the fractional diffusion equation \cite{schneider}. In the presence of an
external potential, it generalises to the time-fractional Fokker-Planck
equation \cite{mebakla,report,report1}, see also below.

(iii) Finally, take $\psi(t)=\delta(t-\tau)$ sharply peaked, but
$\lambda(x)$ of L{\'e}vy stable form with index $0<\mu<2$. The resulting
process is Markovian, but with diverging variance. It can be shown that the
fractional moments scale like \cite{nonn}
\begin{equation}
\label{fmsd}
\langle|x(t)|^{\delta}\rangle\propto \left(K^{(\mu)}t\right)^{\delta/\mu},
\end{equation}
were $K^{(\mu)}=\sigma^{\mu}/\tau$. The upper index $\mu$ is chosen to
distinguish $^{\mu}$ from the subdiffusion constant $K_{\mu}$. Note
that the dimension of $K^{(\mu)}$ is $\mathrm{cm}^{\mu}/\mathrm{sec}$.
From Eq.~(\ref{pfl}) one can immediately obtain the Fourier image of the
associated probability density function,
\begin{equation}
P(k,t)=\exp\Big\{-K^{(\mu)}|k|^{\mu}t\Big\}.
\end{equation}
From Eq.~(\ref{lsd}) this is but a symmetric L{\'e}vy stable density with stable index $\mu$,
and this type of random walk process is most aptly coined a L{\'e}vy flight.
A L{\'e}vy flight manifestly has regular exponential mode relaxation and is in
fact Markovian. However, the modes in position space are no more
sharply localised
like in the Gaussian or subdiffusive case. Instead, individual modes bear
the hallmark of an L{\'e}vy stable density, that is, the diverging variance. We will see below
how the presence of steeper than harmonic external potentials causes a finite
variance of the L{\'e}vy flight, although a power-law form of the probability
density remains.

In the remainder of this paper, we deal with the physical and mathematical
properties of L{\'e}vy flights and subdiffusion.
While mostly we will be concerned with the overdamped case, in the last
section we will address the dynamics in velocity space in the presence of
L{\'e}vy noise, in particular, the question of the diverging variance
of L{\'e}vy flights.

\section{L{\'e}vy flights}

\subsection{Underlying random walk process}

To derive the dynamic equation of a L{\'e}vy flight in the presence of an
external force field $F(x)=-dV(x)/dx$, we pursue two different routes. One
starts with a generalised version of the continuous time random walk,
compare Ref.~\cite{mebakla1} for details; a slightly different derivation
is presented in Ref.~\cite{bamekla}.

To include the local asymmetry of the jump length distribution due to the
force field $F(x)$, we introduce \cite{mebakla1,gme} the generalised transfer
kernel $\Lambda(x,x')=
\lambda(x-x')\left[A(x')\Theta(x-x')+B(x')\Theta(x'-x)\right]$ (and
therefore $\Lambda(x,x')=\Lambda(x';x-x')$). As in standard random walk
theory (compare \cite{weiss}), the coefficients $A(x)$ and $B(x)$
define the local asymmetry for jumping left and right, depending on
the value of $F(x)$. Here, $\Theta(x)$ is the Heaviside jump function.
With the normalisation $\int\Lambda(x',\Delta)d\Delta=1$, the fractional
Fokker-Planck equation (FFPE) ensues \cite{mebakla1}:
\begin{equation}
\label{ffpe}
\frac{\partial}{\partial t}P(x,t)=\left(-\frac{\partial}{\partial x}
\frac{F(x)}{m\eta}+K^{(\mu)}\frac{\partial^{\mu}}{\partial|x|^{\mu}}
\right)P(x,t).
\end{equation}
Remarkably, the presence of the L{\'e}vy stable $\lambda(x)$ only affects
the diffusion term, while the drift term remains unchanged
\cite{fogedby,fogedby1,mebakla1}.
The fractional spatial derivative represents an integrodifferential
operator defined through
\begin{equation}
\frac{\partial^{\mu}}{\partial|x|^{\mu}}P(x,t)=\frac{-1}{2\cos(\pi
\mu/2)\Gamma(2-\mu)}\frac{\partial^2}{\partial x^2}\int_{-\infty}^{
\infty}\frac{P(x',t)}{|x-x'|^{\mu-1}},
\end{equation}
for $1<\mu<2$, and a similar form for $0<\mu<1$
\cite{chechkinjsp,samko,podlubny}.
In Fourier space, for all $0<\mu\le2$ the simple relation
\begin{equation}
\mathscr{F}\left\{\frac{\partial^{\mu}}{\partial|x|^{\mu}}P(x,t)\right\}
=-|k|^{\mu}P(k,t)
\end{equation}
holds. In the Gaussian limit $\mu=2$, all relations above reduce to the
familiar second-order derivatives in $x$ and thus the corresponding $P(x,t)$
is governed by the standard
Fokker-Planck equation.

The FFPE (\ref{ffpe}) can also be derived from the Langevin equation
\cite{fogedby,fogedby1,sune,chechkinjsp}
\begin{equation}
\label{langevin}
\frac{dx(t)}{dt}=-\frac{1}{m\gamma}\frac{dV(x)}{dx}+\xi_{\mu}(t),
\end{equation}
driven by white L{\'e}vy stable noise $\xi_{\mu}(t)$, defined through
$L(\Delta t)=\int_t^{t+\Delta t}\xi_{\mu}\left(t'\right)dt'$ being a
symmetric L{\'e}vy stable density of index $\mu$ with characteristic function $p(k,\Delta t)=
\exp\left(-K_{\mu}|k|^{\mu}\Delta t\right)$ for $0<\mu\le2$.
As with standard Langevin equations, $K^{(\mu)}$ denotes the noise
strength, $m$ is the mass of the diffusing (test) particle, and $\gamma$
is the friction constant characterising the dissipative interaction with
the bath of surrounding particles.

A subtle point about the FFPE (\ref{ffpe}) is that it does not uniquely define
the underlying trajectory \cite{igorsub}; however, starting from our definition
of the process in terms of the stable jump length distribution $\lambda(x)
\sim|x|^{-1-\mu}$, or its generalised pendant $\Lambda(x,x')$, the FFPE
(\ref{ffpe}) truly represents a L{\'e}vy flight in the presence of the force
$F(x)$. This poses certain difficulties when non-trivial boundary conditions
are involved, as shown below.

\subsection{Propagator and symmetries}

\begin{figure}
\includegraphics[width=12cm]{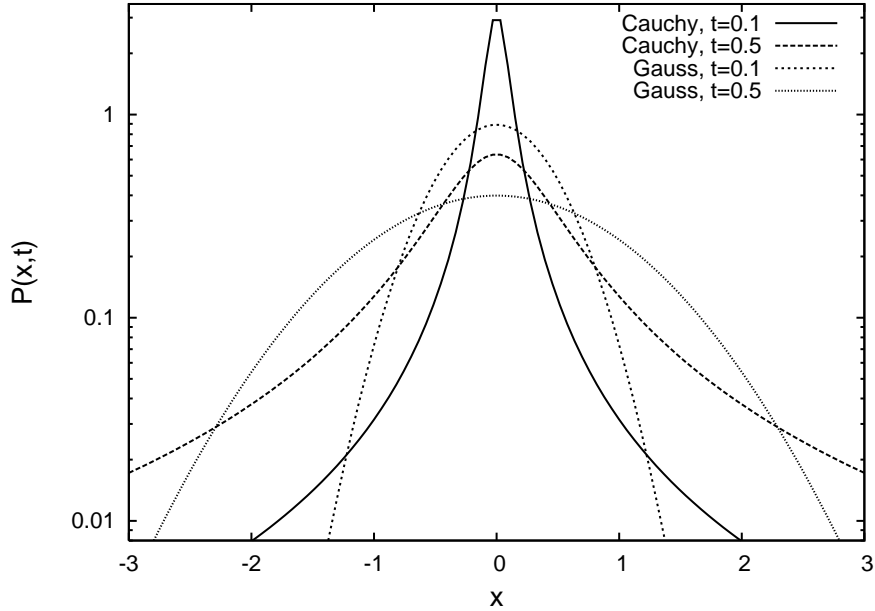}
\caption{Cauchy distribution ($\mu=1$) for two times in comparison to the
Gaussian ($\mu=2$). We chose $K^{(1)}=K=1$. Note that the Cauchy distribution
is narrower at the origin, and after crossing the Gaussian falls off in the
much slower power-law fashion.
\label{cauchy}}
\end{figure}

In absence of an external force, $F(x)=0$, the exact solution of the FFPE is
readily obtained as the L{\'e}vy stable density $P(k,t)=\exp\left(-K^{(\mu)}|k|^{\mu}t\right)$
in Fourier space. Back-transformed to position space, an analytical solution
is given in terms of the Fox $H$-function \cite{report,wegrimeno,sune}
\begin{equation}
P(x,t)=\frac{1}{\mu|x|}H^{1,1}_{2,2}\left[\frac{|x|}{\left(K^{(\mu)}t
\right)^{1/\mu}}\left|\begin{array}{l}(1,1/\mu),(1,1/2)\\(1,1),(1,1/2)
\end{array}\right.\right],
\end{equation}
from which the series expansion
\begin{equation}
P(x,t)=\frac{1}{\mu (K^{(\mu)}t)^{1/\mu}}\sum_{\nu=0}^{\infty}
\frac{\Gamma([1+\nu]/\mu)}{\Gamma([1+\nu]/2)\Gamma(1-[1+\nu]/2)}
\frac{(-1)^{\nu}}{\nu!}\left(\frac{|x|}{(K^{(\mu)}t)^{1/\mu}}\right)^{\nu}
\end{equation}
derives. For $\mu=1$, the propagator reduces to the Cauchy L{\'e}vy stable density
\begin{equation}
P(x,t)=\frac{1}{\pi\left(K^{(1)}t+x^2/[K^{(1)}t]\right)}.
\end{equation}
We plot the time evolution of $P(x,t)$ for the Cauchy case $\mu=1$
in Fig.~\ref{cauchy} in comparison to the limiting Gaussian case $\mu=2$.

\begin{figure}
\includegraphics[width=12cm]{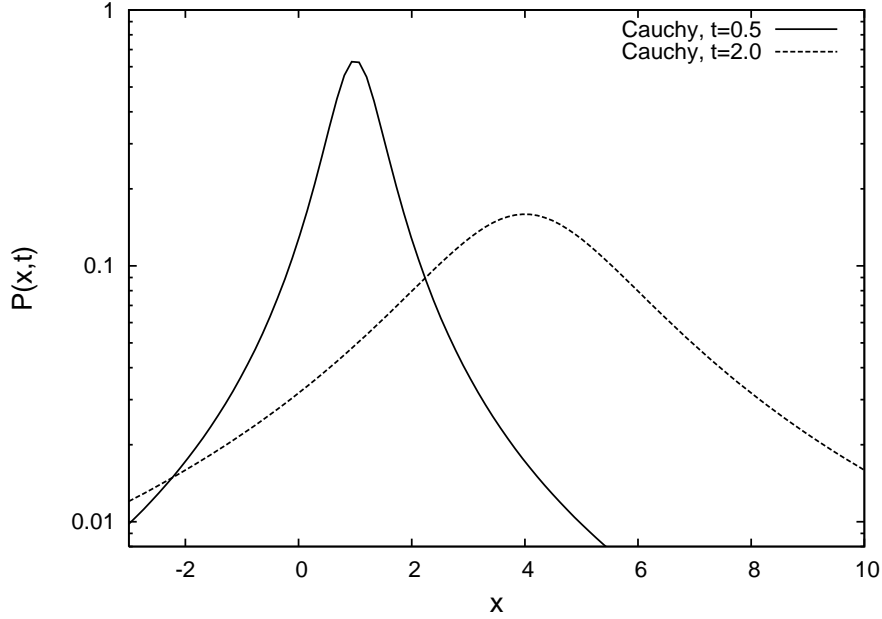}
\caption{Cauchy distribution with $K^{(1)}=1$ advected along a field $F_0/(m
\gamma)=2$, for different times.
\label{drift}}
\end{figure}

Due to the point symmetry of the FFPE (\ref{ffpe}) for $F(x)=0$, the
propagator $P(x,t)$ is invariant under change of sign, and it is monomodal,
i.e., it has its global maximum at $x=0$, the point where the
initial distribution $P(x,0)=\delta(x)$ was launched at $t=0$.
The latter property is lost in the case
of strongly confined L{\'e}vy flights discussed below. Due to their
Markovian character, L{\'e}vy flights also possess a Galilei invariance
\cite{meco,report}. Thus,
under the influence of a constant force field $F(x)=F_0$, the solution of the
FFPE can be expressed in terms of the force-free solution by introducing the
wave variable $x-F_0t$, to obtain
\begin{equation}
\label{gali}
P_{F_0}(x,t)=P_{0}\left(x-\frac{F_0t}{m\gamma},t\right).
\end{equation}
This result follows from the FFPE (\ref{ffpe}), that in Fourier domain becomes
\cite{sune}
\begin{equation}
\frac{\partial}{\partial t}P(k,t)=\left(-ik\frac{F_0}{m\gamma}-K^{(\mu)}
|k|^{\mu}\right)P(k,t),
\end{equation}
with solution
\begin{equation}
P(k,t)=\exp\left(-\left[ik\frac{F_0}{m\gamma}+K^{(\mu)}|k|^{\mu}\right]t
\right).
\end{equation}
By the translation theorem of the Fourier transform, Eq.~(\ref{gali}) yields.
We show an example of the drift superimposed to the dispersional spreading of
the propagator in Fig.~\ref{drift}.

\subsection{Presence of external potentials}

\subsubsection{Harmonic potential}

In an harmonic potential $V(x)=\frac{1}{2}\lambda x^2$, an exact form for the
characteristic function can be found. Thus, from the corresponding FFPE in 
Fourier space,
\begin{equation}
\frac{\partial}{\partial t}P(k,t)=-\frac{\lambda}{m\gamma}k\frac{\partial}{
\partial k}P(k,t)-K^{(\mu)}|k|^{\mu}P(k,t),
\end{equation}
by the method of characteristics one obtains
\begin{equation}
P(k,t)=\exp\left(-\frac{m\gamma K^{(\mu)}|k|^{\mu}}{\lambda\mu}
\left[1-e^{-\mu\lambda t/(m\gamma)}\right]\right)
\end{equation}
for an initially central $\delta$-peak, $P(x,0)=\delta(x)$ \cite{sune}.
This is but the characteristic function of an L{\'e}vy stable density with time-varying width.
For short times, $1-\exp(-\mu\lambda t/[m\gamma])\sim\mu\lambda t/[m
\gamma]$ grows linearly in time, such that $P(k,t)\sim\exp\left(-K^{(\mu)}
|k|^{\mu}t\right)$ as for a free L{\'e}vy flight. At long times, the
stationary solution defined through
\begin{equation}
P_{\mathrm{st}}(k)=\exp\left(-\frac{m\gamma K^{(\mu)}|k|^{\mu}}{\lambda
\mu}\right),
\end{equation}
is reached. Interestingly, it has the same stable index $\mu$ as the
driving L{\'e}vy noise. By separation of variables, a summation formula for
$P(x,t)$ can be obtained similarly to the solution of the Ornstein-Uhlenbeck
process in the presence of white Gaussian noise, however, with the Hermite
polynomials replaced by $H$-functions \cite{sune}.

We note that in the Gaussian limit $\mu=2$, the stationary solution by
necessity has to match the Boltzmann distribution corresponding to $\exp
\left(-k_BTk^2/[2\lambda]\right)$. This requires that the Einstein-Stokes
relation $K=k_BT/(m\gamma)$ is fulfilled \cite{vankampen}. One might
therefore speculate whether for a system driven by external L{\'e}vy noise
a generalised Einstein-Stokes relation should hold, as was established for
the subdiffusive case \cite{mebakla,report}. As will be shown now, in steeper
than harmonic external potentials, the stationary form of $P(x,t)$ even leaves
the basin of attraction of L{\'e}vy stable densities.

\subsubsection{Steeper than harmonic potentials}
\label{multimodal}

To investigate the behaviour of L{\'e}vy flights in potentials, that are
steeper than the harmonic case considered above, we introduce the non-linear
oscillator potential
\begin{equation}
V(x)=\frac{a}{2}x^2+\frac{b}{4}x^4,
\end{equation}
that can be viewed as a next order approximation to a general confining,
symmetric potential. It turns out that the resulting process differs from
above findings if a suitable choice of the ratio $a/b$ is made. For
simplicity, we introduce dimensionless variables through
\begin{equation}
x\to x/x_0;\,\,t\to t/t_0;\,\,a\to at_0/(m\gamma),
\end{equation}
where
\begin{equation}
x_0=\left(\frac{m\gamma K^{(\mu)}}{b}\right)^{1/(2+\mu)};\,\,
t_0=\frac{x_0^{\mu}}{K^{(\mu)}},
\end{equation}
arriving at the FFPE
\begin{equation}
\label{oscill}
\frac{\partial}{\partial t}P(k,t)+|k|^{\mu}=\left(k\frac{\partial^3}{
\partial k^3}-ak\frac{\partial}{\partial k}\right)P(k,t).
\end{equation}

\begin{figure}
\includegraphics[width=12cm]{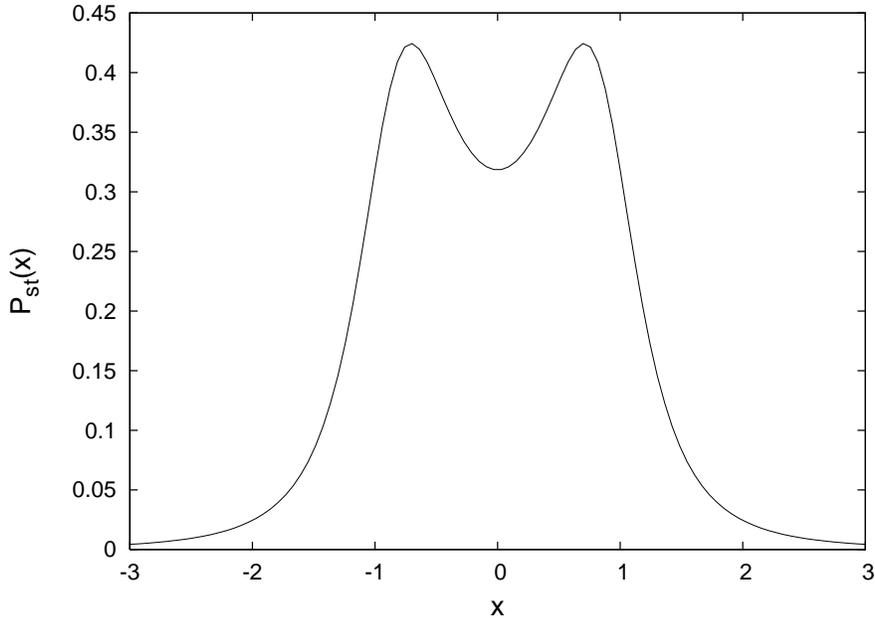}
\caption{Bimodal stationary probability density $P_{\mathrm{st}}(x)$
from Eq.~(\ref{bim}). The maxima are at $\pm\sqrt{1/2}$.
\label{bimod}}
\end{figure}

Consider first the simplest case of a quartic oscillator with $a=0$ in the
presence of Cauchy noise ($\mu=1$). In this limit, the stationary solution
can be obtained exactly, yielding the expression
\begin{equation}
\label{bim}
P_{\mathrm{st}}(x)=\frac{1}{\pi}\frac{1}{1-x^2+x^4}
\end{equation}
plotted in Fig.~\ref{bimod}.
Two distinct new features in comparison to the free L{\'e}vy flight, and the
L{\'e}vy flight in an harmonic potential: (1) Instead of the maximum at $x=0$,
one observes two maxima positioned at
\begin{equation}
x_m=\pm\sqrt{1/2};
\end{equation}
at $x=0$, we find a local minimum. (2) There occurs a power-law asymptote
\begin{equation}
P_{\mathrm{st}}(x)\sim\frac{1}{\pi x^4}
\end{equation}
for $x\gg 1$; consequently, this stationary solution no longer represents
an L{\'e}vy stable density, and the associated mean squared displacement is finite, $\langle
x^2\rangle<\infty$.

A more detailed analysis of Eq.~(\ref{oscill}) reveals
\cite{chechkinpre,chechkinjsp}, that (i) the bimodality
of $P(x,t)$ occurs only if the amplitude of the harmonic term, $a$, is below
a critical value $a_c$; (ii) for general $\mu$, the asymptotic behaviour
is $P_{\mathrm{st}}(x)\sim\pi^{-1}\sin(\pi\mu/2)\Gamma(\mu)|x|^{-\mu
-3}$; (iii) and there exists a finite bifurcation time $t_c$ at which the
initially monomodal form of $P(x,t)$ acquires a zero curvature at $x=0$,
before settling in the terminal bimodal form.

Interestingly, in the more general power-law behaviour
\begin{equation}
V(x)=\frac{|x|^c}{c},
\end{equation}
the turnover from monomodal to bimodal form of $P(x,t)$ occurs exactly when
$c>2$. The harmonic potential is therefore a limiting case when the solution
of the FFPE still belongs to the class of L{\'e}vy stable densities and follows the generalised
central limit theorem. This is broken in a superharmonic (steeper than
harmonic) potential. The corresponding bifurcation time $t_c$ is finite
for all $c>2$ \cite{chechkinjsp}. An additional effect appears when $c>4$:
there exists a transient trimodal state when the relaxing $\delta(x)$-peak
overlaps with the forming humps at $x=\pm x_m$. At the same time, the variance
is finite, if only $c>4-\mu$, following from the asymptotic stationary
solution
\begin{equation}
P_{\mathrm{st}}(x)\sim\frac{\sin(\pi\mu/2)\Gamma(\mu)}{\pi|x|^{
\mu+c-1}}.
\end{equation}
Details of the asymptotic behaviour and the bifurcations can be found
in Refs.~\cite{chechkinpre,chechkinjsp}. From a reverse engineering
point of view, L{\'e}vy flights in confining potentials are studied in
\cite{iddo}.

\subsection{First passage and first arrival of L{\'e}vy flights}

One might naively expect that a jump process of L{\'e}vy type, whose
variance diverges (unless confined in a steep potential) may lead to
ambiguities when boundary conditions are introduced, such as an 
absorbing boundary at finite $x$. Indeed, it is conceivable that for a
jump process with extremely long jumps, it becomes ambiguous how to
properly define the boundary condition: should the test particle be
absorbed when it arrives exactly \emph{at\/} the boundary, or when
it crosses it \emph{anyplace\/} during a non-local jump?

This question is trivial in the case of a narrow jump length distribution:
all steps are small, and the particle cannot jump across a point (in the
continuum limit considered herein). For such processes, one enforces
a Cauchy boundary condition $P(0,t)=0$ at the point $x=0$ of the absorbing
boundary, removing the particle once it hits the barrier after starting at
$x_0$, where the dynamics is governed by Eq.~(\ref{ffpe}) with $F(x)=0$.
Its solution can easily be obtained by standard methods, for instance, the
method of images. This is completely equivalent to considering the \emph{first
arrival\/} to the point $x=0$, expressed in terms of the diffusion equation
with sink term:
\begin{equation}
\label{sink}
\frac{\partial}{\partial t}\mathscr{P}(x,t)=K\frac{\partial^2}{\partial x^2}
\mathscr{P}(x,t)-p_{\mathrm{
fa}}(t)\delta(x),
\end{equation}
defined such that $P(0,t)=0$. Note that the quantity $\mathscr{P}$ is no 
longer a probability density, as probability decays to zero; for this
reason, we use the notation $\mathscr{P}$. From Eq.~(\ref{sink}) by
integration we obtain the survival probability
\begin{equation}
\mathscr{S}(t)=\int\mathscr{P}(x,t)dx
\end{equation}
with $\mathscr{S}(0)=1$ and $\lim_{t\to\infty}\mathscr{S}(t)=0$. Then, the
first arrival density becomes
\begin{equation}
p_{\mathrm{fa}}(t)=-\frac{d}{dt}\mathscr{S}(t).
\end{equation}
Eq.~(\ref{sink}) can be solved by standard methods (determining the homogeneous
and inhomogeneous solutions). It is then possible to express $\mathscr{P}(x,t)$
in terms of the propagator $P(x,t)$, the solution of Eq.~(\ref{ffpe}) with
$F(x)=0$ with the same initial condition, $P(x,0)=\delta(x-x_0)$ and natural
boundary conditions. One obtains
\begin{equation}
P(0,t)=\int_0^tp_{\mathrm{fa}}(\tau)P(x_0,t-\tau)d\tau,
\end{equation}
such that the first arrival density corresponds to the waiting time distribution
to jump from $x_0$ to 0 (or, vice versa, since the problem is symmetric). In
Laplace space, this relation takes on the simple algebraic form $p_{\mathrm{fa}}
(u)=P(0,u)/P(x_0,u)$. Both methods the explicit boundary value problem and the
first arrival problem for Gaussian processes produce the well-known first
passage (or arrival) density of L{\'e}vy-Smirnov type (\ref{smirnov}),
\begin{equation}
\label{gfpt}
p(t)=p_{\mathrm{fa}}(t)=\frac{x_0}{\sqrt{4\pi Kt^3}}\exp\left(-\frac{x_0^2}{
4Kt}\right)\sim\frac{x_0}{\sqrt{4\pi Kt^3}},
\end{equation}
with the asymptotic power-law decay $p(t)\sim t^{-3/2}$, such that no mean
first passage time exists \cite{hughes,risken}.

Long-tailed jump length distributions of L{\'e}vy stable form, however, endow
the test particle with the possibility to jump across a certain point
repeatedly. The first arrival necessarily becomes less efficient. Indeed, as
shown in Ref.~\cite{chechkinbvp}, the Gaussian result (\ref{gfpt}) is
generalised to
\begin{equation}
\label{lfpt}
p_{\mathrm{fa}}(t)\sim C(\mu)\frac{x_0^{\mu-1}}{\left(K^{(\mu)}\right)
^{1-1/\mu}t^{1-1/\mu}}, \,\,\, \mbox{as } t\to\infty
\end{equation}
with $C(\mu)=\mu\Gamma(2-\mu)\Gamma(2-1/\mu)\sin(\pi[2-\mu]/2)
\sin^2(\pi/\mu)/(\pi^2[\mu-1])$, and $1<\mu\ge 2$ \cite{chechkinbvp}.
The long-time decay $\sim t^{-2+1/\mu}$ is slower than in (\ref{gfpt}).



\begin{figure}
\includegraphics[width=12cm]{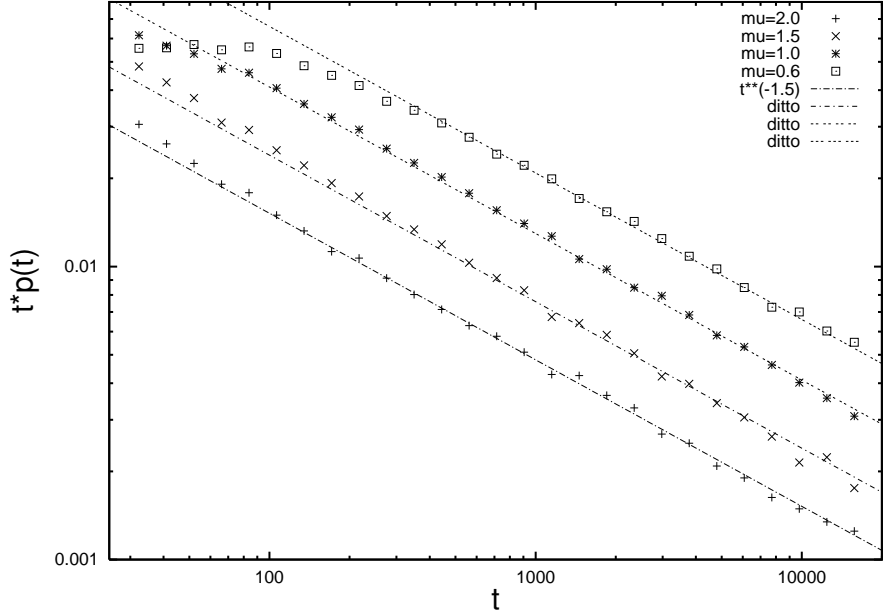}
\caption{First passage density for various stable indices $\mu$ and initial
position $x_0=10.0$ away from the absorbing boundary. Again, the universal
$\sim t^{-3/2}$ scaling is distinct. Note that we plot $tp(t)$ in the $y$-axis.
\label{fpt_1}}
\end{figure}

One might naively assume that the first passage problem (the particle is
removed once it crosses the boundary) for L{\'e}vy flights
should be more efficient, that is, the first passage density $p(t)$ should
decay quicker, than for a narrow jump length distribution. However, as we
have a symmetric jump length distribution $\lambda(x)$, the long outliers
characteristic for these L{\'e}vy flights can occur both toward and away from
the absorbing barrier. From this point of view it is not totally surprising
to see the simulations result in Fig.~\ref{fpt_1}, that clearly
indicate a universal asymptotic decay $\sim t^{-3/2}$, exactly as for the
Gaussian case.


In fact, for all Markovian processes with a \emph{symmetric\/} jump length
distribution, the Sparre Andersen theorem \cite{sparre,sparre1,feller,redner} proves
without knowing any details about $\lambda(x)$ the asymptotic behaviour of
the first passage time density universally follows $p(t)\sim t^{-3/2}$. The
details of the specific form of $\lambda(x)$ only enter the prefactor, and
the pre-asymptotic behaviour. A special case of the Sparre Andersen theorem
was proved in Ref.~\cite{frisch} when the particle is released at $x_0=0$ at
time $t=0$, and after the first jump an absorbing boundary is installed at
$x=0$. This latter case was simulated extensively in Ref.~\cite{zukla}. From
a fractional diffusion equation point of view, it was shown in Ref.~\cite{chechkinbvp}
that the fractional operator $\partial^{\mu}/\partial|x|^{\mu}$ needs to
be modified, to account for the fact that $\mathscr{P}(x,t)\equiv0$ beyond the
absorbing boundary, such that long-range correlations are present exclusively
for all $x$ in the semi-axis containing $x_0$. The fractional diffusion
equation in the presence of the absorbing boundary therefore has to be
modified to \cite{chechkinbvp}
\begin{equation}
\label{wienhopf}
\frac{\partial}{\partial}\mathscr{P}(x,t)=\frac{K_{\mu}}{\kappa}\frac{
\partial^2}{\partial x^2}\int_0^{\infty}\frac{\mathscr{P}(x',t)}{|x-x'|^{
\mu-1}}dx',
\end{equation}
where $\kappa=2\Gamma(2-\mu)\left|\cos(\pi\mu/2)\right|$,
such that the first term on the right hand side no longer represents a
Fourier convolution. An approximate solution with Cauchy boundary condition
reveals $p(u)\sim1-cu^{1/2}$, where $c$ is a constant, indeed leading to the
Sparre Andersen behaviour $p(t)\sim t^{-3/2}$.

This also demonstrates that the method of images no longer applies when
L{\'e}vy flights are considered, for the images
solution
\begin{equation}
\mathscr{P}_{\mathrm{im}}(x,t)=P(x-x_0,t)-P(x+x_0,t)
\end{equation}
would be governed by the \emph{full\/} fractional diffusion equation, and not
Eq.~(\ref{wienhopf}), and the result for the first passage density, $p(t)\sim
t^{-1-1/\mu}$ would decay faster than the Sparre Andersen universal
behaviour. A detailed discussion of the applicability of the method of images
is given in terms of a subordination argument in Ref.~\cite{igorsub}. We
emphasise that this subtle failure of the method of images has been overlooked
in literature previously \cite{west,gittermann}, and care should therefore
be taken when working with results based on such derivations. We also note
that the method of images works in cases of subdiffusion, as the step length
is narrow \cite{meklabvp}.

\subsection{Leapover properties of L{\'e}vy flights}

\begin{figure}
\includegraphics[width=12cm]{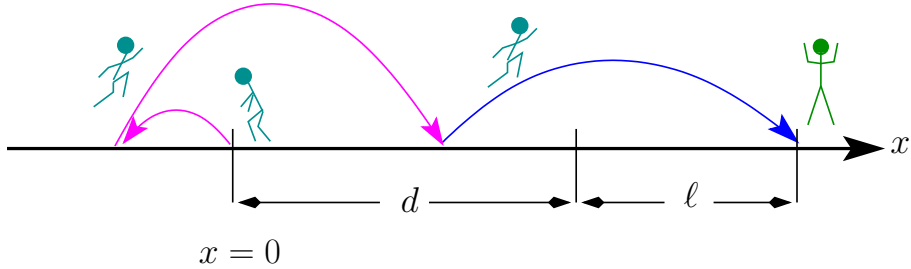}
\caption{Schematic of the leapover problem: the random walker starts at
$x=0$ and after a number of jumps crosses the point $x=d$, overshooting
it by a distance $\ell$. For narrow jump length distributions, each jump
is so small that crossing of the point $d$ is equal to arriving at this
point.}
\label{fig0}
\end{figure}

The statistics of first passage times is a classical concept to quantify
processes in which it is of interest when the dynamic variable crosses
a certain threshold value for the first time. For processes with broad
jump length distributions, another quantity is of interest, namely, the
statistics of the first passage leapovers, that is, the distance
the random walker overshoots the threshold value $x=d$ in a single jump
(see Fig.~\ref{fig0}). Surprisingly, for symmetric LFs with jump length
distribution $\lambda(x)\sim|x|^{-1-\mu}$ ($0<\mu<2$) the distribution
of leapover lengths across $x_0$ is distributed like $p_l(\ell)\sim\ell^
{-1-\mu/2}$, i.e., it is much broader than the original jump length
distribution. In contrast, for one-sided LFs jumps the scaling of $p_l
(\ell)$ bears the same index $\mu$. Information on the leapover behaviour
is important to the understanding of how far search processes of animals for
food or of proteins for their specific binding site along DNA
overshoot their target, or to define better stock market
strategies determining when to buy or sell a certain share instead of a
given threshold price.

Using a general theorem for the first passage times and leapovers for
homogeneous processes with independent increments, leapover properties
for completely symmetric and
fully asymmetric (one-sided) LFs are derived in \cite{leapover}, the general
case is considered in \cite{leapover1}. The basic
results are as follows. For the completely symmetric LF with index $\mu$,
the distribution of first passage leapover lengths $\ell$ for a particle
originally released a distance $d$ away from the boundary reads (in scaled
units)
\begin{equation}
p_l(\ell)=\frac{\sin(\pi\mu/2)}{\pi}\frac{d^{\mu/2}}{\ell^{\mu/2}(d+
\ell)}.
\end{equation}
The validity of this result is confirmed by
extensive simulations, for more details refer to
\cite{leapover,leapover1,kochekla}.
Note that $p_l$ is normalised. In the limit $\mu\to2$, $p_l$ tends
to zero if $\ell\neq0$ and to infinity at $\ell=0$ corresponding to the
absence of leapovers in the Gaussian continuum limit. However, for $0<\mu
<2$ the leapover PDF follows an asymptotic power-law with index $\mu/2$,
and is thus broader than the original jump length PDF $\lambda(x)$ with index
$\mu$. This is a remarkable finding: while $\lambda$ for $1<\mu<2$ has
a finite characteristic length $\langle|x|\rangle$, this always diverges for
$p_l(\ell)$ irrespective of $\mu$.


In contrast, the result for completely asymmetric LFs has the form
\cite{iddo,leapover}
\begin{equation}
p_l(q)=\langle e^{-q\ell}\rangle=\frac{\sin(\pi\mu)}{\pi}\int_0^{\infty}
e^{-q\ell}\frac{x^{\mu}}{\ell^{\mu}(x+\ell)},
\end{equation}
leading to the leapover PDF
\begin{equation}
\label{lod}
p_l(\ell)=\frac{\sin(\pi\mu)}{\pi}\frac{d^{\mu}}{\ell^{\mu}(d+\ell)},
\end{equation}
which corresponds to the result obtained in Refs.~\cite{iddoyossi,tal} from a
different method. Thus, for the one-sided LF, the scaling of the leapover
is exactly the same as for the jump length distribution, namely, with exponent
$\mu$. Again, this result compares favourably with simulations
\cite{leapover,leapover1}.


\subsection{Kramers problem for L{\'e}vy flights}

Many physical and chemical problems are related to the thermal fluctuations
driven crossing of an energetic barrier, such as dissociation of molecules,
nucleation processes, or the escape from an external, confining potential of
finite height \cite{haenggi}. A particular example of barrier crossing in a
double-well potential driven by L{\'e}vy noise was proposed for a long time
series of paleoclimatic data \cite{ditlevsen}. Further cases where the
crossing of a potential barrier driven by L{\'e}vy noise is of interest is
in the theory of plasma devices \cite{chechkin}, among others \cite{report1}.

\begin{figure}
\includegraphics[width=12cm]{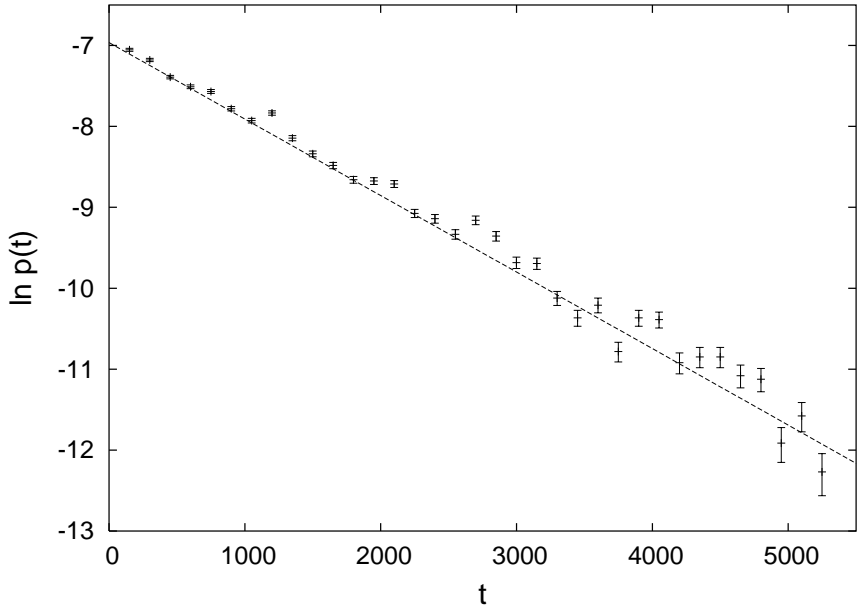}
\caption{Probability density function $p(t)$ of barrier crossing times for
$\mu=1.0$ and $D=10^{-2.5}\approx 0.00316$. The dashed line is a fit to
equation (\protect\ref{decay}) with MCT $T_c=1057.8\pm 17.7$.
\label{expo}}
\end{figure}

To investigate the detailed behaviour of barrier crossing under the influence
of external L{\'e}vy noise, we choose the rather generic double well shape
\begin{equation}
V(x)=-\frac{a}{2}x^2+\frac{b}{4}x^4.
\end{equation}
Integrating the Langevin equation (\ref{langevin}) with white L{\'e}vy
noise,
we find an exponential decay of the survival density in the initial well:
\begin{equation}
\label{decay}
p(t)=\frac{1}{T_c}\exp\left(-\frac{t}{T_c}\right),
\end{equation}
as demonstrated in Fig.~\ref{expo}. L{\'e}vy flight processes being Markovian,
this is not surprising, since the mode relaxation is exponential
\cite{report,report1}. More interesting is the question how the mean
escape time $T_c$ behaves as function of the characteristic noise parameters
$D$ and $\mu$. While in the regular Kramers problem with Gaussian driving
noise the Arrhenius-type activation $T_c=C\exp(h/D)$ is followed, where $h$
is the barrier height, and the prefactor $C$ includes details of the potential,
in the case of L{\'e}vy noise, a power-law form
\begin{equation}
T_c(\mu,D)=\frac{C(\mu)}{\left(K^{(\mu)}\right)^{\mu(\mu)}}
\end{equation}
was assumed \cite{levykramers}. Detailed investigations \cite{levykramers1}
show that the
scaling exponent $\mu(\mu)=1$ for all $\mu$ strictly smaller than 2.
As already proposed in Ref.~\cite{ditlevsen1} and derived in \cite{imkeller}
in a somewhat different model, this means that, apart from a prefactor, the
L{\'e}vy flight is insensitive to the external potential for the barrier
crossing, as confirmed by simulations \cite{levykramers1}.
Note that in comparison
to Ref.~\cite{levykramers}, also values of $\mu$ in the range $(0,1)$ are
included. For large values of $D$, deviations from the scaling are observed:
eventually it will only take a single jump to cross the barrier when $D\to
\infty$. Detailed studies show indeed that eventually the unit time step is
reached, i.e., $T_c\to 1$.

\subsection{More on the "pathology"}

Despite their mathematical foundation due to the generalised central limit
theorem and their broad use in the sciences and beyond as description for
statistical quantities, and despite the existence of systems (for instance,
the diffusion on a polymer in chemical space mediated by jumps where the
polymer loops back on itself \cite{igor,dirk,michael}), the divergence of
the fluctuations of L{\'e}vy processes is sometimes considered a pathology.
This was already put forward by West and Seshadri \cite{wese}, who pointed
out that a L{\'e}vy flight in velocity space would be equivalent to a
diverging kinetic energy. Here, we show that higher order dissipation
effects lead to natural cutoffs in L{\'e}vy processes.

At higher velocities the friction experienced by a moving body starts to
depend on the velocity itself \cite{bogol}. Such non-linear friction
is known from the classical Riccati equation $Mdv(t)
/dt=Mg-Kv(t)^2$ for the fall of a particle of mass $M$ in a gravitational
field with acceleration $g$ \cite{davis}, or autonomous oscillatory systems
with a friction that is non-linear in the velocity \cite{bogol,andronov}. The
occurrence of a non-constant friction coefficient $\gamma(V)$ leading to a
non-linear dissipative force $-\gamma(V)V$ was highlighted in Klimontovich's
theory of non-linear Brownian motion \cite{klimontovich}. It is therefore
natural that higher order, non-linear friction terms also occur in the case
of L{\'e}vy processes.

We consider the velocity-dependent dissipative non-linear form (necessarily
an even function) \cite{dnl}
\begin{equation}
\label{dnl}
\gamma(V)=\gamma_0+\gamma_2V^2+\ldots+\gamma_{2N}V^{2N} \quad
\therefore \gamma_{2N}>0
\end{equation}
for the friction coefficient of the L{\'e}vy flight in velocity space as
governed by the Langevin equation
\begin{equation}
\label{vlangevin}
dV(t)+\gamma(V) V(t)dt=dL(t)
\end{equation}
with the constant friction $\gamma_0=\gamma(0)$. $L(t)$ is the $\mu$-stable
L{\'e}vy
noise defined in terms of a characteristic function $p^*(\omega,t)=
\mathscr{F}\{p(L,t)\}\equiv\int_{-\infty}^{\infty}p(L,t)\exp\left(i\omega
L\right)dL$ of the form
$p^*(\omega,t)=\exp\left(-D|\omega|^{\mu}t\right)$
\cite{levy,samo,uchaikin},
where $D$ of dimension ${\rm cm}^{\mu}/{\rm sec}$ is the generalised
diffusion constant. This is equivalent to the fractional
Fokker Planck equation
\cite{fogedby,fogedby1,chechkin,chechkinjsp,report,report1}
\begin{equation}
\label{vfpe}
\frac{\partial P(V,t)}{\partial t}=\frac{\partial}{\partial V}
\Big(V\gamma(V)P\Big)+D\frac{\partial^{\mu}P}{\partial |V|^{\mu}}.
\end{equation}

\begin{figure}
\includegraphics[width=12cm]{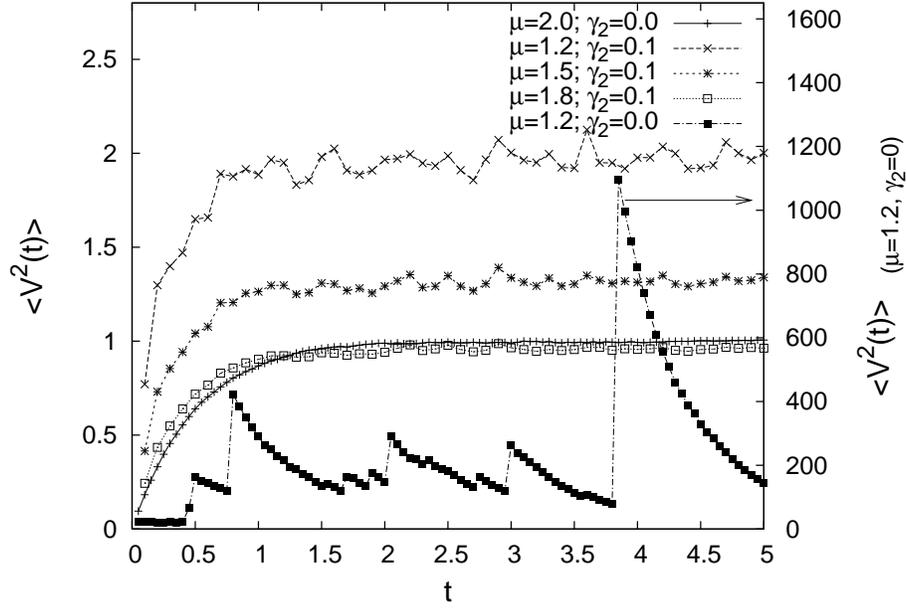}
\caption{Variance $\langle V^2(t)\rangle$ as function of time $t$, with the
quartic term set to zero, $\gamma_4=0$ and $\gamma_0=1.0$ for all cases. The
variance is finite for the cases $\mu=2.0,\gamma_2=0.0$; $\mu=1.2,
\gamma_2=0.1$; $\mu=1.5,\gamma_2=0.1$; and $\mu=1.8,\gamma_2=0.1$. These
correspond to the left ordinate. For the case $\mu=1.2,\gamma_2=0.0$, the
variance diverges, strong fluctuations are visible; note the large values
of this curve corresponding to the right ordinate.
\label{dissnonlin}}
\end{figure}

As we showed in Sec.~\ref{multimodal} by the example of the L{\'e}vy flights
in position space, the presence of the first higher
order correction, $\gamma_2V^2$ in the friction coefficient $\gamma(V)$
rectifies the L{\'e}vy motion such that the asymptotic power-law becomes
steeper and the variance finite. When even higher order corrections are
taken into consideration, also higher order moments become finite. We
show an example in Fig.~\ref{dissnonlin} for the second moment.

\begin{figure}
\includegraphics[width=12cm]{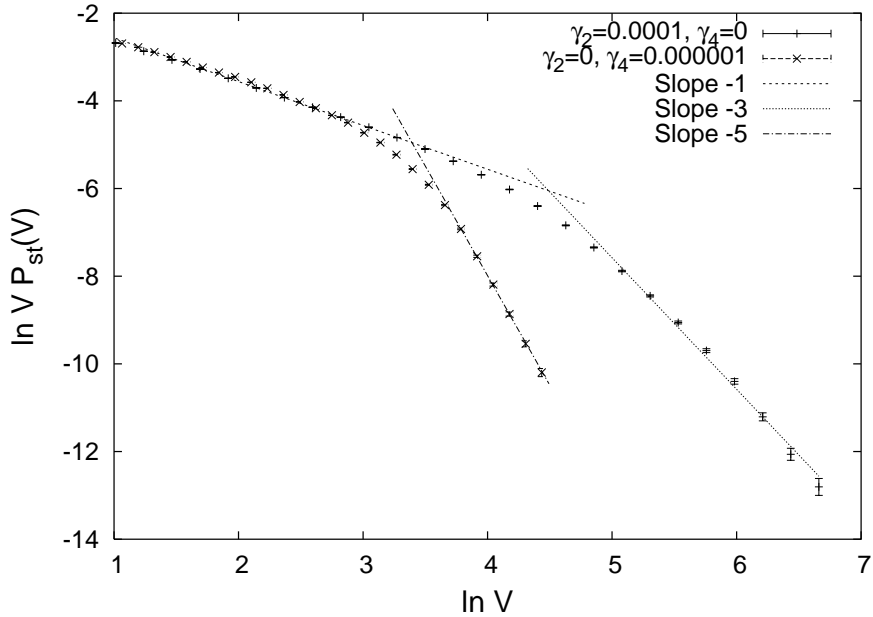}
\caption{Stationary PDF $P_{\mathrm{st}}(V)$ for $\gamma_0=1.0$ and (i)
$\gamma_2=0.0001$ and $\gamma_4=0$; and (ii) $\gamma_2=0$ and $\gamma_4=
0.000001$; with $\mu=1.0$.
The lines indicate the slopes $-1$, $-3$, and $-5$.
\label{dissnonlin1}}
\end{figure}

The effect on the velocity distribution of the process defined by
Eqs.~(\ref{vlangevin}) and (\ref{vfpe}) for higher order corrections are
demonstrated in Fig.~\ref{dissnonlin1} for the stationary limit, $P_{\mathrm{
st}}(V)=\lim_{t\to\infty}P(V,t)$: while for smaller $V$ the character of the
original L{\'e}vy stable behaviour is preserved (the original power-law
behaviour, that is, persists to intermediately large $V$), for even larger $V$
the corrections due to the dissipative non-linearity are visible in the
transition(s) to steeper slope(s).

These dissipative non-linearities remove the divergence of the kinetic energy
from the measurable subsystem of the random walker. In the ideal mathematical
language, the surrounding bath provides an infinite amount of energy through
the L{\'e}vy noise, and the coupling via the non-linear friction dissipates
an infinite amount of energy into the bath, and thereby introduces a natural
cutoff in the kinetic energy distribution of the random walker subsystem.
Physically, such divergencies are not expected, but correspond to the limiting
procedure of large numbers in probability theory. We showed
that both statements can be reconciled, and that L{\'e}vy processes are
indeed physical.

Also Gaussian continuum diffusion exhibits non-physical features, possibly the
most prominent being the infinite propagation speed inherent of the parabolic
nature of the diffusion equation: even at very short times after system
preparation in, say, a state $P(x,0)=\delta(x)$, there has already arrived
a finite portion of probability at large $x$. This problem can be corrected
by changing from the diffusion to the Cattaneo (telegrapher's) equation.
Still, for most purposes, the uncorrected diffusion equation is used.
Similarly, one often uses natural boundary conditions even though the
system under consideration is finite, since one might not be interested in
the behaviour at times when a significant portion of probability has reached
the boundaries. In a similar sense, we showed that "somewhere out in the 
wings" L{\'e}vy flights are naturally cut off by dissipative non-linear
effects. However, instead of introducing artificial cutoffs, knowing that
for all purposes L{\'e}vy flights are a good quantitative description and
therefore meaningful, we use "pure" L{\'e}vy stable laws in physical models.

\subsection{Bi-fractional transport equations}

The coexistence of long-tailed forms for both jump length and waiting time
PDFs was investigated within the CTRW approach in Ref.~\cite{zuklaa},
discussing in detail the laminar-localised phases in chaotic dynamics. In
the framework of fractional transport equations, the combination of the
waiting time PDF $\psi(t)\sim t^{-1-\alpha}$ ($0<\alpha<1$) and jump length
PDF $\lambda(x)\sim|x|^{-1-\mu}$ leads to a dynamical equation with fractional
derivatives in respect to both time and space
\cite{luchko,mainardi,nonn,weno}:
\begin{equation}
\label{bifde}
\frac{\partial}{\partial t}P(x,t)=K_{\alpha}^{(\mu)}\, _0D_t^{1-\alpha}\frac{
\partial^{\mu}}{\partial |x|^{\mu}}P(x,t).
\end{equation}
As long as the condition $1\le\alpha\le\mu\le 2$ is met, both exponents can
be chosen within the entire range \cite{mainardi}. For $\mu=2$, in particular,
this equation covers both sub- and superdiffusion up to ballistic motion,
the latter corresponding to the wave equation \cite{abm}.
A closed form solution of Eq.~(\ref{bifde}) can be found in terms of Fox's
$H$-functions, see Ref.~\cite{nonn}, where also some special cases permitting
elementary solutions are considered.
Bi-fractional diffusion equations were also discussed in
Refs.~\cite{barkaicp,saichev1,baeumer,hughes1,gorenflo2,uchaikin2}.
A bi-fractional Fokker-Planck equation with a power-law dependence
$\propto |x|^{-\theta}$ ($\theta\in\mathbb{R}$) of the diffusion coefficient
was studied in Refs.~\cite{fa,lenzi}.

\subsection{L{\'e}vy walks}

L{\'e}vy walks correspond to the spatiotemporally coupled version
of continuous time random walks. The waiting time and jump length PDFs,
that is, are no longer decoupled but appear as conditional in the
form $\psi(x,t)\equiv \lambda(x)p(t|x)$ (or $\psi(t)\tilde{p}(x|t)$)
\cite{klablushle}. In particular, through the coupling $p(t|x)=\frac{1}{2}
\delta\left(|x|-vt^{\nu}\right)$, one introduces a generalised velocity
$v$, which penalises long jumps such that the overall process, the L{\'e}vy
walk, attains a finite variance and a PDF with two spiky fronts successively
exploring space \cite{zuklan,zuklan1}. Thus, L{\'e}vy walks have similar
properties to generalised Cattaneo/telegraphers' equation-type models
\cite{come,meco,meno}. As we here focus on the properties of transport
processes governed by the Langevin equation under L{\'e}vy noise, we
only briefly introduce L{\'e}vy walks.

On the basis of fractional equations, formulations were obtained for the
description of L{\'e}vy walks in the presence of non-trivial
external force fields, with the same restriction to lower order moments
in respect to an L{\'e}vy walk process \cite{basil,meso}. Recently, however, a
coupled fractional equations was reported \cite{some}, which describes
a force-free L{\'e}vy walk exactly. Thus, it was shown that the fractional
version
of the material derivative $\partial/\partial t\pm\partial/\partial x$,
\begin{equation}
\label{fmd}
d_{\pm}^{\beta}P(x,t)\equiv \, _0D_t^{\beta}P(x\pm t,t),
\end{equation}
defined in Fourier-Laplace space through
\begin{equation}
\mathscr{F}\left\{\mathscr{L}\left\{d_{\pm}^{\beta}f(x,t);u\right\};k
\right\}\equiv (u\pm ik)^{\beta}f(k,u)
\end{equation}
($\mathscr{F}$ acts on $x$ and $\mathscr{L}$ on $t$) replaces the uncoupled
fractional
time operators, see also the detailed discussion of L{\'e}vy walk processes
in Ref.~\cite{zuklan}. Although one may argue
for certain forms \cite{some}, there is so far no derivation for the
incorporation of general external force fields in the coupled formalism.
We note that a very similar fractional approach to L{\'e}vy walks was
suggested in Ref.~\cite{meerschaert}.

\section{Subdiffusion and the fractional Fokker-Planck equation}

\subsection{Physical foundation of subdiffusion}

The stochastic motion of a Brownian particle of mass $m$ is described by the
Langevin equation \cite{langevin,vankampen,chandrasekhar}
\begin{equation}
m\frac{d^2x}{dt^2}=-m\eta_1v+F(x)+m\mathit{\Gamma}(t),\,\,v=\frac{dx}{dt},
\end{equation}
where $F(x)$ is an external force field, and $\eta_1$ is the friction
coefficient. The erroneous bombardment through
the surrounding bath molecules is described by the fluctuating noise $\mathit{
\Gamma}(t)$. To properly describe Brownian motion, $\mathit{\Gamma}(t)$ has to
be chosen $\delta$-correlated (white) and Gaussian distributed. This is, the
time averages of
$\mathit{\Gamma}(t)$ are: $\overline{\mathit{\Gamma}(t)}=0$, and $\overline{
\mathit{\Gamma}(t)\mathit{\Gamma}(t')}=D\delta(t-t')$, where $D$ is the noise
strength. After averaging of
the fluctuations, the velocity moments become \cite{chandrasekhar}
\begin{equation}
\langle\Delta v\rangle=\left(\eta v-\frac{F(x)}{m}\right)\Delta t,\,\,
\langle(\Delta v)^2\rangle=\frac{2\eta k_BT}{m}\Delta t+\mathcal{O}\left(
[\Delta t]^2\right).
\end{equation}
Both are proportional to $\Delta t$. For regular Brownian motion, these
increments are used as expansion coefficients in the Chapman-Kolmogorov
equation \cite{chandrasekhar}.

Subdiffusion now comes about through a so-called trapping scenario. Trapping
describes the occasional immobilisation of the test particle for a waiting
time distributed according to the distribution $\psi(t)$. Here we assume
that the particle leaves the trap with the same velocity it had prior to
immobilisation (this condition can be relaxed). In between trapping
events, the particle is assumed to follow the regular Langevin equation such
that each motion event on average lasts for a mean time $\tau^*$. Choosing
$\psi(t)$ with a finite characteristic waiting time one can show that this
trapping scenario indeed preserves Brownian motion.
However, once the characteristic waiting time diverges, it can
be shown by application of the generalised central limit theorem that the
occurrence of a large number of trapping events leads to subdiffusion
and the fractional Klein-Kramers equation for the joint PDF $P(x,v,t)$
\cite{gcke,meklacke,meklacke1}. Integrating out the velocity coordinate,
the fractional Fokker-Planck equation emerges for the PDF $P(x,t)$:
\begin{equation}
\label{ffp}
\frac{\partial P(x,t)}{\partial t}=\,_0D_t^{1-\alpha}\frac{\partial}{\partial
x}\left(-\frac{F(x)}{m\eta_{\alpha}}+K_{\alpha}\frac{\partial}{\partial x}
\right)P(x,t).
\end{equation}

In mathematical terms the trapping scenario corresponds to a subordination
principle: Trapping events cause a transformation of the internal step time
of the random walk to a different observation time, corresponding to the
relation
\begin{equation}
\label{subord}
P(x,t)=\int_0^{\infty}\mathscr{E}_{\alpha}(s,t)P_1(x,s)ds,
\end{equation}
where $P_1(x,t)$ is the solution of the regular Brownian Fokker-Planck
equation with $\alpha=1$. Explicit forms of the kernel $\mathscr{E}_{\alpha}
(s,t)$ are known in terms of Fox' $H$-function \cite{report}. In the Laplace
domain, the kernel $\mathscr{E}_{\alpha}(s,t)$ has the comparatively simple
form
\begin{equation}
\mathscr{E}_{\alpha}(s,u)
\frac{\eta_{\alpha}}{\eta_1u^{1-\alpha}}\exp\left(-\frac{\eta_{\alpha}}{\eta_1}
u^{\alpha}s\right).
\end{equation}
Note that this transformation guarantees the existence and positivity of the
solution $P(x,t)$ of the fractional Fokker-Planck equation if only the
corresponding Brownian problem possesses a well-defined solution.
We note that recently an alternative approach to continuous time random walk
subdiffusion in terms of $\delta$-noise spike trains was presented
\cite{saichev}.

\subsection{Linear response and fluctuation-dissipation relation}

From the fractional Fokker-Planck equation (\ref{ffp}) it is straightforward
to prove that in the presence of a constant field $F_0$ the linear response
relation
\begin{equation}
\label{linresp}
\langle x(t)\rangle_{F_0}=\frac{k_BT}{2}\langle x^2(t)\rangle_{F=0}
\end{equation}
is fulfilled \cite{mebakla,report}. This is due to the fact that the waiting
time distribution $\psi(t)$ is independent of the force. Given the
comparatively large traps required to create the long-tailed form of $\psi(t)$,
this assumption appears reasonable for not too large $F_0$. Experimentally,
the linear response relation (\ref{linresp}) was verified \cite{schiff}.

The stationary solution of the fractional Fokker-Planck equation (\ref{ffp})
is the standard Boltzmann-Gibbs equilibrium, $\lim_{t\to\infty}P(x,t)=\mathscr{
N}\exp\left\{-V(x)/[k_BT]\right\}$, where $\mathscr{N}$ is a normalisation
constant \cite{mebakla}. This result is immediately obvious from the picture
of subordination, that changes the temporal spacing of events but preserves
the causal mode relaxation. As $F(x)=-V'(x)$, from the stationary solution
of Eq.~(\ref{ffp}) by comparison with the Boltzmann-Gibbs form one recovers
the generalised Einstein-Stokes relation \cite{mebakla}
\begin{equation}
K_{\alpha}=\frac{k_BT}{m\eta_{\alpha}},
\end{equation}
reflecting the preservation of the fluctuation-dissipation relation for the
subdiffusion process. An experimental verification of this relation was
reported by \cite{amblard}.

\subsection{Propagator}

\begin{figure}
\includegraphics[width=8cm]{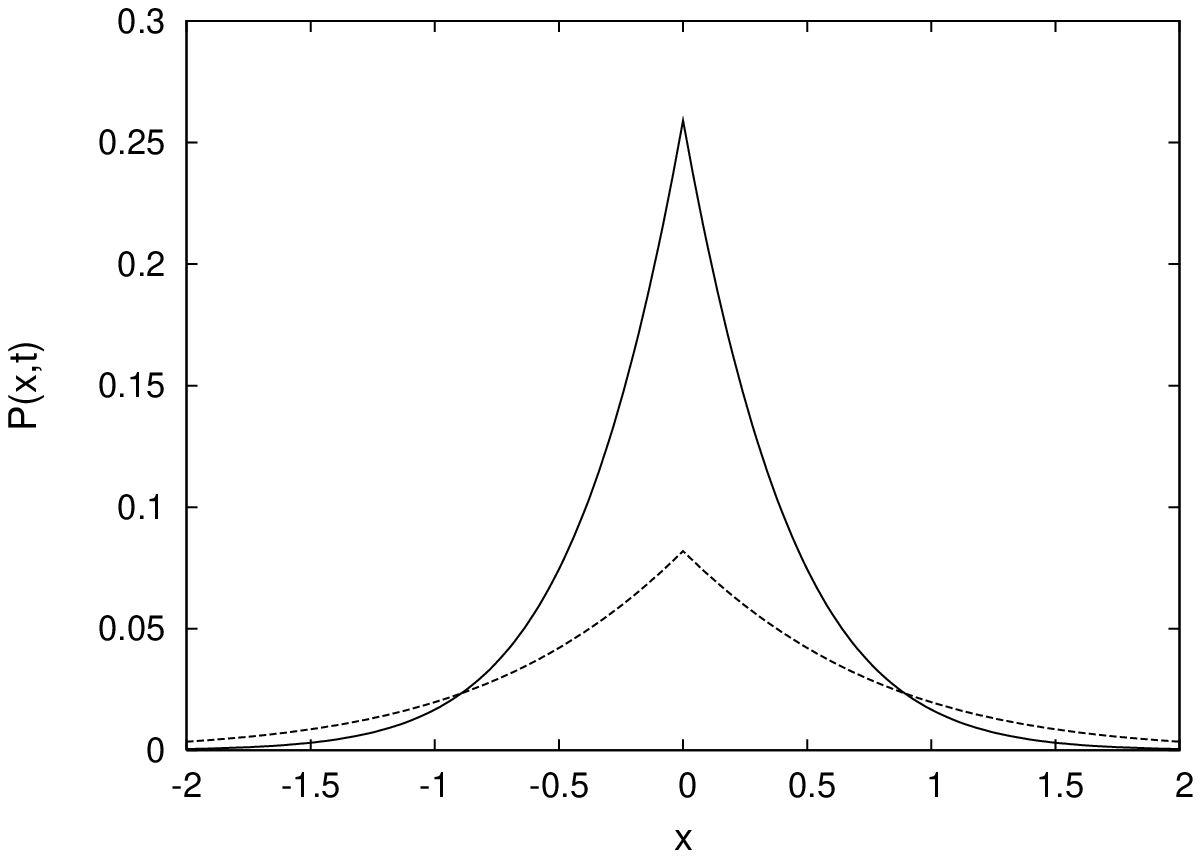}
\includegraphics[width=8cm]{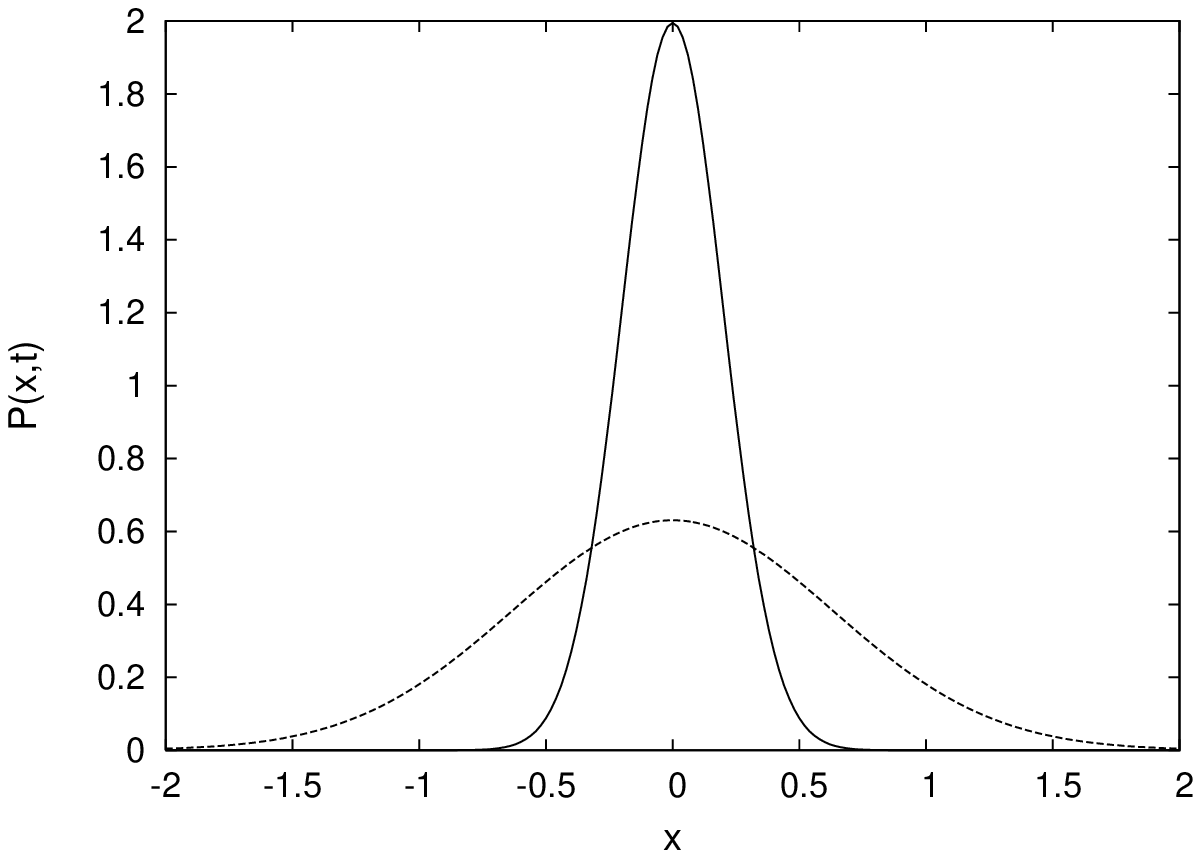}
\caption{Propagator $P(x,t)$ of the fractional diffusion equation with
$\alpha=1/2$ (left) and the normal diffusion equation (right) for consecutive
times.}
\label{fig5}
\end{figure}

In absence of an external force, the fractional diffusion
equation is solved by Fox' $H$-function \cite{report},
\begin{equation}
\label{fdsol1}
W(x,t)=\frac{1}{\sqrt{4K_{\alpha}t^{\alpha}}}H^{1,0}_{1,1}\left[
\frac{|x|}{\sqrt{K_{\alpha}t^{\alpha}}} \left| \begin{array}{l}
\left(1-\frac{\alpha}{2},\frac{\alpha}{2}\right)\\
(0,1) \end{array} \right. \right],
\end{equation}
an equivalent form to the original solution by \cite{schneider}. Its
series expansion reads \cite{report}
\begin{equation}
W(x,t)=\frac{1}{\sqrt{4K_{\alpha}t^{\alpha}}}\sum_{n=0}^{\infty}
\frac{(-1)^n}{n!\Gamma(1-\alpha[n+1]/2)}\left(\frac{x^2}{K_{\alpha}
t^{\alpha}}\right)^{n/2},
\end{equation}
and asymptotically reaches the stretched Gaussian form
\begin{eqnarray}
\nonumber
W(x,t)&\sim& \frac{1}{\sqrt{4\pi K_{\alpha}t^{\alpha}}} \sqrt{\frac{1}{
2-\alpha}} \left(\frac{2}{\alpha}\right)^{(1-\alpha)/(2-\alpha)}
\left(\frac{|x|}{\sqrt{K_{\alpha}t^{\alpha}}}\right)^{-
(1-\alpha)/(2-\alpha)}\\
&&\times \exp\left(- \frac{2-\alpha}{2}
\left(\frac{\alpha}{2}\right)^{\alpha/(2-\alpha)}
\left[\frac{|x|}{\sqrt{K_{\alpha}t^{\alpha}}}\right]^{1/(1-\alpha/2)}\right).
\label{sg}
\end{eqnarray}
valid for $|x|\gg \sqrt{K_{\alpha}t^{\alpha}}$. The latter corresponds to the
known result from continuous time random walk theory \cite{zuklan,zuklan1}.
The $H$-function simplifies if the exponent $\alpha$ is a rational number.
For instance, for $\alpha=1/2$, it can be rewritten
in terms of the Meijer $G$-function
\begin{equation}
P(x,t)=\frac{1}{\sqrt{8\pi^3 K_{1/2}t^{1/2}}}G^{3,0}_{0,3}\left[
\left(\frac{x^2}{16K_{1/2}t^{1/2}}\right)^2\left| \begin{array}{l}
\rule[0.06in]{0.4in}{0.01in}
\\ 0,\frac{1}{4},\frac{1}{2} \end{array}\right. \right],
\end{equation}
that is known in symbolic mathematics packages such as Mathematica.
Using this representation in Fig.~\ref{fig5} we show the subdiffusive
propagator with its pronounced cusps at the site of the initial condition,
in comparison to the smooth Gaussian propagator of Brownian motion.

\subsection{Boundary value and first passage time problems}

In the presence of reflecting, absorbing, or mixed, boundary conditions, the
narrow jump length distribution of subdiffusive processes makes it possible
to use analogous techniques to calculate the propagator as known from normal
diffusion. Thus, separation into eigenmodes or the method of images can be
applied, the only difference entering through the time-dependence of the modes.
A very convenient way is to employ the Brownian result for a given geometry,
and subordinate that result to obtain the behaviour for a subdiffusing
particle, according to Eq.~(\ref{subord}).

In the presence of an absorbing boundary an important quantity characterising
the dynamics of the system is the survival probability
\begin{equation}
\mathscr{S}_{\alpha}(t)=\int_{\mathbb{D}}\mathscr{P}(x,t)dx,
\end{equation}
where $\mathbb{D}$ denotes the interval over which $\mathscr{P}(x,t)$ is
defined. The
initial value of the survival probability is $\mathscr{S}_{\alpha}(0)=1$,
and it decays to zero for long times. From $\mathscr{S}_{\alpha}(t)$, we
can define the first passage time density
\begin{equation}
\label{fptd}
p_{\alpha}(t)=-\frac{d}{dt}\mathscr{S}_{\alpha}(t).
\end{equation}
The mean first passage time is
\begin{equation}
\mathcal{T}=\int_0^{\infty}p_{\alpha}(t)tdt.
\end{equation}
Note that from Eqs.~(\ref{fptd}) and (\ref{subord}), we obtain the relation
\begin{equation}
p_{\alpha}(u)=p_1\left(u^{\alpha}\right)
\end{equation}
between the subdiffusive and Brownian results in Laplace space. Thus, these
are connected by a relation similar to the subordination (\ref{subord}),
with kernel $\exp\left(-su^{\alpha}\right)$ instead of $\mathscr{E}_{\alpha}
(s,u)$ \cite{report1,yossibj}.

For the three most prominent cases of first passage time problems, we obtain
the following subdiffusive generalisations:

(i) For subdiffusion in the semi-infinite domain with an absorbing wall at the
origin and initial condition $P(x,0)=\delta(x-x_0)$ it was found that
\cite{meklabvp}
\begin{equation}
p(t)\sim\frac{x_0}{|\Gamma(-\alpha/2)|K_{\alpha}^{1/2}}t^{-1-\alpha/2},
\end{equation}
i.e., the decay becomes a flatter power-law than in the Markovian case
where $p(t)\sim t^{-3/2}$.

(ii) Subdiffusion in the semi-infinite domain in the presence of an external
bias $V$ falls off faster, but still in power-law manner
\cite{barkai,yossibj,grl}:
\begin{equation}
p(t)\sim t^{-1-\alpha}.
\end{equation}
In strong contrast to the biased Brownian case, we now end up with a process
whose characteristic time scale diverges. This is exactly the mirror of the
multiple trapping model, i.e., the classical motion events become repeatedly
interrupted such that the immobilisation time dominates the process.
In contrast, for $\alpha=1$ the result
\begin{equation}
p_1(t)=\frac{x_0}{\sqrt{4\pi K_{\alpha}t^3}}\exp\left(-\frac{(x_0-Vt)^2}{4K
t}\right)
\end{equation}
is valid, producing the classical form $\mathcal{T}=x_0/V$ for the mean first
passage time.

(iii) Subdiffusion in a finite box \cite{meklabvp}:
\begin{equation}
p(t)\sim t^{-1-\alpha},
\end{equation}
i.e., this process leads to the same scaling behaviour for longer times as
found for the biased semi-infinite case (ii).

The latter two results should be compared to the classical Scher-Montroll
finding for the first passage time density of biased motion in a finite
system of size $L$ with absorbing boundary condition. In that case, the
first passage time density exhibits two power-laws
\begin{equation}
p(t)\sim\left\{\begin{array}{ll}
t^{\alpha-1}, & t<\tau\\
t^{-1-\alpha}, & t>\tau
\end{array}\right.
\end{equation}
the sum of whose exponents equals $-2$
\cite{pfister,pfister,schermo,grl}. Here,
$\tau$ is a system size dependent time scale \cite{pfister}.

\subsection{Fractional Ornstein-Uhlenbeck process}

\begin{figure}
\includegraphics[width=12cm]{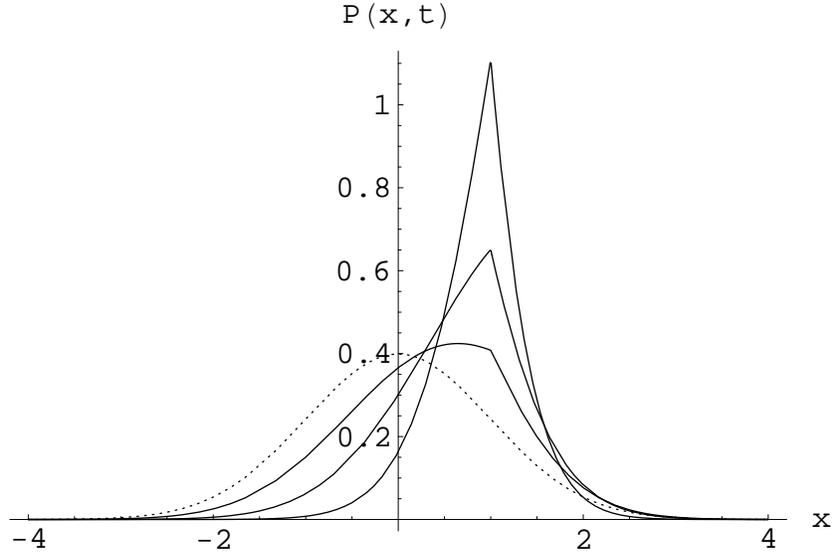}
\caption{Time evolution of the PDF of the fractional Ornstein-Uhlenbeck process
($\alpha=1/2$). The initial condition was chosen as $\delta(x-1)$. Note the
strongly persistent cusp at the location of the initial peak. Dimensionless
times: 0.02, 0.2, 2. The dashed line corresponds to the Boltzmann equilibrium.
\label{fig1}}
\end{figure}

The Ornstein-Uhlenbeck process corresponds to the motion in an harmonic
potential $V(x)=\frac{1}{2}m\omega^2x^2$
giving rise to the restoring force field $F(x)=-m\omega^2x$, i.e., to the
dynamical equation
\begin{equation}
\frac{\partial}{\partial t}P(x,t)=\,_0D_t^{1-\alpha}\left(\frac{\partial}{
\partial x}
\frac{\omega^2x}{\eta_{\alpha}}+K_{\alpha}\frac{\partial^2}{\partial x^2}
\right)P(x,t).
\end{equation}
From
separation of variables, and the definition of the Hermite polynomials
\cite{abramowitz}, one finds the series solution for the fractional
Fokker-Planck equation with the Ornstein-Uhlenbeck potential
\cite{mebakla,report},
\begin{eqnarray}
\nonumber
P(x,t)=\sqrt{\frac{m\omega^2}{2\pi k_BT}}&&\sum_{n=0}^{\infty}\frac{1}{2^nn!}
E_{\alpha}\left(-\frac{n\omega^2t^{\alpha}}{\eta_{\alpha}}\right)H_n\left(
\frac{\sqrt{m}\omega x_0}{\sqrt{2 k_BT}}\right)\\
&&\times H_n\left(\frac{\sqrt{m}\omega
x}{\sqrt{2 k_BT}}\right)\exp\left(-\frac{m\omega^2x^2}{2k_BT}\right)
\label{herm}
\end{eqnarray}
plotted in Fig.~\ref{fig1}.
Individual spatial eigenmodes follow the ordinary Hermite polynomials of
increasing order, while their temporal relaxation is of Mittag-Leffler form,
with decreasing internal time scale $\left(\eta_{\alpha}/[n\omega^2]\right)^{
1/\alpha}$. Numerically, the solution (\ref{herm}) is somewhat cumbersome
to treat. In order to plot the PDF $P(x,t)$ in Fig.~\ref{fig1}, it is
preferable to use the closed form solution (we use dimensionless variables)
\begin{equation}
\label{ou_closed}
P(x,t)=\frac{1}{\sqrt{2\pi\left(1-e^{-2t}\right)}}
\exp\left(-\frac{\left(x-x_0e^{-t}\right)^2}{2\left(1-e^{-2t}
\right)}\right)
\end{equation}
of the Brownian case, and the transformation (\ref{subord}) to construct the
fractional analogue.

\begin{figure}
\includegraphics[width=12cm]{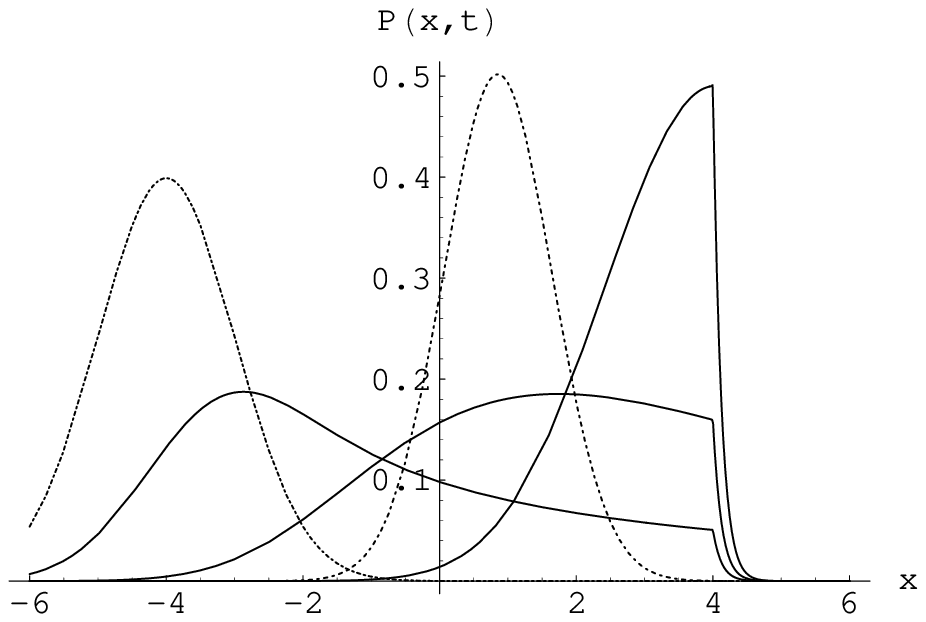}
\caption{Time evolution of the PDF of the fractional Ornstein-Uhlenbeck
process with superposed constant force of dimensionless strength $V=-4$
($\alpha=1/2$). The initial condition was chosen as $\delta(x-4)$.
Dimensionless times: 0.02, 0.2, 2. The dashed lines corresponds to the
Brownian solution at times 0.5 and 50 (in essence, the stationary state).
Again, note the cusps due to the initial condition, causing a strongly
asymmetric shape of the PDF in contrast to the Gaussian nature of the
Brownian counterpart.
\label{ouforce}}
\end{figure}
\begin{figure}
\includegraphics[width=12cm]{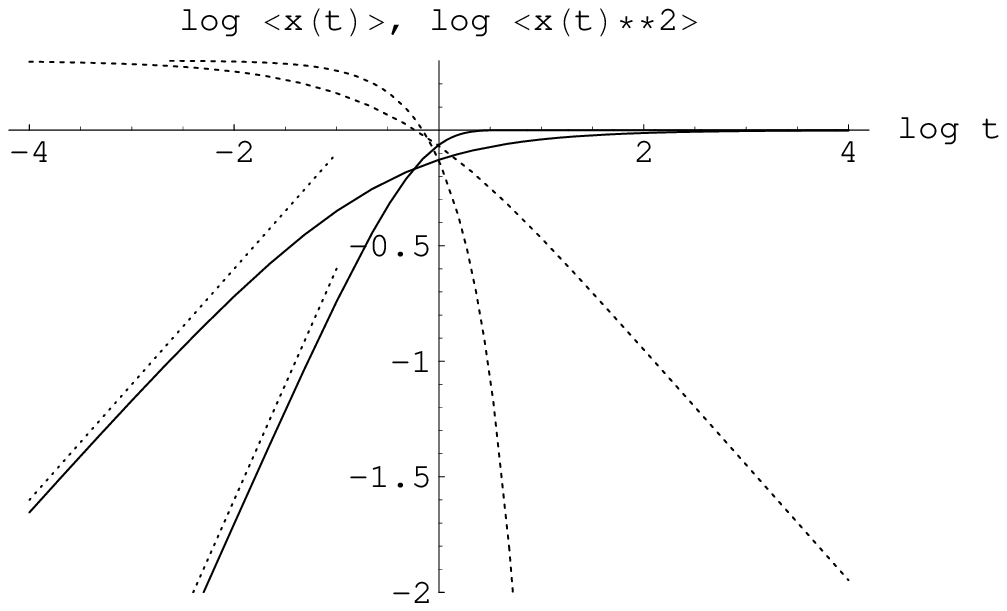}
\caption{First ($x_0=2$, dashed line) and second ($x_0=0$, full line) moment
of the fractional Ornstein-Uhlenbeck process ($\alpha=1/2$), in comparison to
the Brownian case. $\log_{10}$$-$$\log_{10}$ scale. The dotted straight lines
show the initial (sub)diffusive behaviour with slopes 1/2 and 1, in the
special case $x_0=0$ chosen for the second moment.
\label{oumoments}}
\end{figure}

Fig.~\ref{fig1} shows the distinct cusps at the position of the initial
condition at $x_0=1$.
The relaxation to the final Gaussian Gibbs-Boltzmann PDF can be seen from the
sequence of three consecutive times. Only at stationarity, the cusp gives way
to the smooth Gaussian shape of the equilibrium PDF. By adding an
additional linear drift $V$ to the harmonic restoring force, the drift term
in the FFPE (\ref{ffp}) changes to $-\partial (x-V)P(x,t)/\partial x$, and
the exponential in expression (\ref{ou_closed}) takes on the form $\exp\left(
-\left[x-V-(x_0-V)e^{-t}\right]/\left[2\left(1-e^{-2t}\right)\right]\right)$.
As displayed in Fig.~\ref{ouforce}, the strong persistence of the initial
condition causes a highly asymmetric shape of the PDF, whereas the Brownian
solution shown in dashed lines retains its symmetric Gaussian profile.

Let us finally address the moments of the fractional Ornstein-Uhlenbeck
process, Eq.~(\ref{herm}).
These can be readily obtained either from the Brownian result
with the integral transformation (\ref{subord}), or from integration $\int
dx x^n \cdot$ of the FFPE (\ref{ffp}). For the first and second moments
one obtains:
\begin{equation}
\langle x(t)\rangle=x_0E_{\alpha}\left(-\frac{\omega^2t^{\alpha}}{\eta_{
\alpha}}\right)
\end{equation}
and
\begin{equation}
\langle x(t)^2\rangle=x_{\rm th}^2+\left(x_0^2-x_{\rm th}^2\right)E_{\alpha}
\left(-\frac{2\omega^2t^{\alpha}}{\eta_{\alpha}}\right),
\end{equation}
respectively. The first moment starts off at the initial position, $x_0$,
and then falls off in a Mittag-Leffler pattern, reaching the terminal inverse
power-law $\sim t^{-\alpha}$. The second moment turns from the initial value
$x_0^2$ to the thermal value $x_{\rm th}^2=k_BT/(m\omega^2)$. In the special
case $x_0=0$, the second moment measures initial force-free diffusion due
to the initial exploration of the flat apex of the potential. We graph the two
moments in Fig.~\ref{oumoments} in comparison to their Brownian counterparts.

\section{Future directions}

Anomalous diffusion is becoming widely recognised in a variety of fields.
Apart from the anomalous spreading of tracers and the consequences for
the propagator, additional questions such as ageing and weak ergodicity
breaking become important for the understanding of experiments and their
modelling on complex systems. At the same time, the theory of anomalous
processes is expanding. For instance, regarding questions on the weak
ergodicity breaking,
the calculation of multipoint moments or the correct introduction
of cutoffs are currently being worked on, to complete the world
\emph{Beyond Brownian Motion\/} \cite{physicstoday}.

\section*{Acknowledgements}

RM acknowledges partial financial support through the Natural Sciences and
Engineering Research Council (NSERC) of Canada and the Canada Research Chairs
program of the Government of Canada. AVC acknowledges partial support from
the Deutsche Forschungsgemeinschaft (DFG).

\clearpage

\section{Bibliography}

\textbf{Books and Reviews}


\begin{thebibliography}{99}
\bibitem{gorenflo2}
Fractional diffusion: probability distributions and random walk models.
Physica A-Statistical Mechanics and its Applications 305: 106-112
\end{thebibliography}
\end{document}